\documentclass[prd,aps,a4paper,nofootinbib,eqsecnum,twocolumn]{revtex4}  

\newif\ifusesec
\usesectrue  
   
\usepackage{graphicx} 
\usepackage{mathrsfs}
\usepackage{amsmath,amsfonts,amssymb}
\usepackage{multirow}

\newcommand{\beq}{\begin{equation}}
\newcommand{\eeq}{\end{equation}}
\newcommand{\bea}{\begin{eqnarray}}
\newcommand{\eea}{\end{eqnarray}}
\newcommand{\fl}{}

\begin{document}

\title{Particle motion in a rotating dust spacetime}

\author{Davide Astesiano$^{1}$, Donato Bini$^{2,3}$, Andrea Geralico$^{2}$, Matteo Luca Ruggiero$^{4,5}$}
\affiliation{
$^1$\ 
Science Institute, University of Iceland, Dunhaga 3, 107, Reykjavik, Iceland
}
\affiliation{
$^2$\
Istituto per le Applicazioni del Calcolo  M. Picone,  CNR, I-00185 Rome, Italy
}
\affiliation{
$^3$\
INFN, Sezione di Roma Tre, I-00146 Rome, Italy
}
\affiliation{
$^4$\
Department of Mathematics, University of Turin, I-10124 Torino, Italy
}
\affiliation{
$^5$\
INFN - LNL, Viale dell'Universit\'a  2, 35020 Legnaro (PD), Italy\\
}

\begin{abstract}
We investigate the geometrical properties, spectral classification, geodesics, and causal structure of the Bonnor's spacetime [Journal of Physics A Math. Gen., \textbf{10},  1673  (1977)], i.e., a stationary axisymmetric solution with a rotating dust as a source.  
This spacetime has a directional singularity at the origin of the coordinates (related to the diverging vorticity field of the fluid there), which is surrounded by a toroidal region where closed timelike curves (CTCs) are allowed, leading to chronology violations. 
We use the effective potential approach to provide a classification of the different kind of geodesic orbits on the symmetry plane as well as to study the helical-like motion aroud the symmetry axis on a cylinder with constant radius. 
In the former case we find that as a general feature for positive values of the angular momentum test particles released from a fixed space point and directed towards the singularity are repelled and scattered back as soon as they approach the CTC boundary, without reaching the central singularity.
In contrast, for negative values of the angular momentum there exist conditions in the parameter space for which particles are allowed to enter the pathological region. Finally, as a more realistic mechanism, we study accelerated orbits undergoing friction forces due to the interaction with the background fluid, which may also act in order to prevent particles from approaching the CTC region.
\end{abstract}

%\pacno{04.20.Cv}

\date{\today}

\maketitle

\section{Introduction}

General relativistic fluid solutions are typically used to build stellar models, whenever the fluid is distributed within a bounded spacetime region. 
In order to obtain a physical model Einstein's field equations are then coupled to the thermodynamical equations, which are solved all together by imposing several conditions, e.g., symmetries, equation of state, the fall-off behavior of energy density and pressure and their values at the boundary. A further problem in this case is the matching with an asymptotically flat exterior vacuum solution.
If the fluid is filling the whole spacetime, instead, the corresponding solution to the gravitational field equations can be used as a cosmological model. 
There exists very few classes of exact solutions \cite{Stephani:2003tm}, most of them being associated with perfect fluids. For instance, according to the Friedmann equations underlying the standard cosmological model the matter-dominated and radiation-dominated epochs are driven by dust fluids and radiation fields, respectively.
The simplest solutions for the stellar interiors are instead those which are static and spherically symmetric, which can be always matched across the spherical boundary to an external solution, which consists in a Schwarzschild spacetime. 
Our motivation here is to contribute to a deeper understanding of such a kind of solutions, which hold potential relevance across various astrophysical contexts, e.g., in relation to galactic dynamics \cite{Cooperstock:2006dt,Cooperstock:2005qw,Ilyas:2016nio,Balasin:2006cg,Astesiano:2022ozl,Astesiano:2022ghr,Ruggiero:2023ker,Ruggiero:2023geh}.

Unfortunately, most of the known exact solutions are plagued by the presence of singularities as well as unphysical spacetime regions often associated with chronology violations. This is the reason why they have been poorly investigated since their discovery.
The most popular is perhaps the G\"odel spacetime \cite{Godel:1949ga}, which is the prototype for rotating cosmological models.
The source of the gravitational field is represented by a dust of particles which are at rest with respect to the coordinates, but form a family of twisting world lines preserving the cylindrical symmetry of the spacetime.
G\"odel solution admits the existence of closed timelike curves (CTCs), allowing for time travel, so that it has been discarded as describing an unphysical universe.

The aim of the present paper is to study the general properties of the Bonnor's rotating dust cloud spacetime \cite{Bonnor:1977}, belonging to the Van Stockum class of stationary and axisymmetric perfect fluid solutions \cite{Stephani:2003tm}, which are referred to as rigidly rotating dust spacetimes.
In fact, the particles of the dust are free falling along the integral curves of the temporal Killing vector. Their angular velocity with respect to locally nonrotating observers equals in magnitude the angular velocity of dragging of inertial frames.
The dust particles thus form a geodesic congruence with vanishing expansion and shear, but nonzero vorticity. 
Bonnor solution depends on a single parameter, the vanishing of which gives back the flat Minkowskian spacetime. Such a special feature allows for disentangling at any moment genuine curved spacetime properties from special relativistic ones. 
The arbitrary parameter has the meaning of a rotation parameter, since the metric at spatial infinity reduces to that of a spinning body  situated at the origin with angular momentum proportional to it, but with zero mass.
Remarkably, for this solution the density gradient has a component parallel to the rotation axis, which is a peculiar general relativistic feature, as discussed by Bonnor \cite{Bonnor:1977}. He also conjectured the presence of a negative mass distribution contained in the singularity at the center, balancing the positive mass outside, since the density is everywhere positive.
The nature of the central singularity in Bonnor spacetime and the possibility to remove it have been further investigated by many authors \cite{Bonnor:1980wm,deAraujo:2000cs,Bonnor:2005,Bratek:2006uw,Bonnor:2008,Gurlebeck:2009vrw}.
Furthermore, in recent years there has been a renewed interest in this solution (and in other rotating fluid spacetimes) as an alternative to supermassive black holes at the centers of galaxies (see, e.g., Ref. \cite{Ilyas:2016nio}).

The existence of CTCs has also been extensively studied over the years \cite{Bonnor:1980wm,Steadman:1999,Collas:2004hjm,Collas:2004ydx,Lindsay:2016xpc}. Not much attention has been devoted instead to a systematic study of particles motion in Bonnor spacetime. 
A careful investigation of null geodesic motion can be found in Ref. \cite{Steadman:1999}. It has been shown there that there exists a region, or regions, around the origin and the axis of rotation which cannot be entered by null geodesics from spatial infinity. Noticeably, for a wide range of values of the angular momentum photons can be confined in a closed region whose inner boundary is the CTC surface.

In this work we first review the main geometrical properties of the Bonnor spacetime, including the behavior of curvature invariants and the directional character of the central singularity, the spectral classification, and causal structure.
We then focus on timelike geodesics, and on how the presence of both the singularity and the CTC surface affects the motion of test particles. The different kind of orbits are classified by using the effective potential approach. 
In the case of motion on the symmetry plane, some features are familiar from test particle dynamics around compact objects, e.g., circular motion, bound orbits between a minimum and a maximum radius, scattering orbits.
We also discuss conditions on the orbital as well as background parameters such that particles may enter the region containing CTCs.
A more realistic model taking into account the interaction with the dust particles would introduce conditions which may prevent particles to enter the CTC region.
To this aim we introduce a friction force (modeled \'a la Poynting-Robertson \cite{poy1,poy2}, for example) modifying the geodesic behavior, the effect of which is mainly that of slowing-down motion towards the singularity and even stop particles before they reach the pathological region.

We use a mostly positive signature of the metric and units such that $c=1=G$.
Greek indices rum from 0 to 3 while Latin indices from 1 to 3.

\section{Van Stockum class of perfect fluid solutions}

Let us consider a spacetime sourced by a dust fluid, i.e., with energy-momentum tensor $T^{\mu\nu}=\rho_m u^\mu u^\nu$,
associated with a  metric written in cylindrical-like coordinates $x^\alpha\equiv (t,r,\phi,z)$ in the form
\bea\label{IR2} 
    ds^2&=& ds^2_{(t,\phi)}+ds^2_{(r,z)}\,,
\eea
where
\bea
ds^2_{(t,\phi)}&=& g_{tt}dt^2+2g_{t\phi}dtd\phi +g_{\phi\phi}d\phi^2\,,\nonumber\\
ds^2_{(r,z)}&=& g_{rr}dr^2+g_{zz}dz^2\,,
\eea
belonging to the Van Stockum class.
All metric and fluid functions depend on $r$ and $z$, and are chosen so that $g_{rr}=g_{zz}=e^\psi$, and
\bea
g_{tt}&=& -\left(H\gamma^2-\frac{r^2\chi^2}{H\gamma^2}\right) \,,\nonumber\\ 
g_{t\phi}&=& -\frac{r^2\chi}{H\gamma^2} \,,\nonumber\\ 
g_{\phi\phi}&=&\frac{r^2}{H\gamma^2}\,, 
\eea
where we introduced $ \gamma=(1-v^2)^{-1/2}$, as a function of $v$, whose meaning will be clarified below, and  $g_{t\phi}/g_{\phi\phi}\equiv -\chi$.
The  fluid particles move along circular orbits, with four velocity  
\beq
u=\frac{1}{\sqrt{H}}(\partial_t+\Omega \partial_\phi)\,,
\eeq
where the angular velocity $\Omega$ is given by
\beq
\Omega = \chi \pm \frac{H\gamma^2 v}{r} \,,
\eeq
fixed by the (timelike) normalization condition of $u$, $u~\cdot~u~=-1$. Summarizing: the  four functions $H, \chi, \psi, \gamma$ are used to parametrize the metric tensor, and are related by the Einstein's equations, which are rather involved in this general case, and will not be displayed (see, e.g., Ref. \cite{Stephani:2003tm}).

It is also convenient to introduce the lapse-shift notation and write the metric in a form adapted to the Zero-Angular-Momentum-Observers (ZAMOs), with four-velocity\footnote{The symbols $\flat$ and $\sharp$ are associated with the fully covariant and fully contravariant representation of tensors, respectively.}
\beq
n^\flat=-N dt\,,\qquad n^\sharp=\frac{1}{N}\left(\partial_t -N^a\partial_a \right)\,,
\eeq
namely
\beq
N=\gamma \sqrt{H} \,,\qquad N^r=0\,,\qquad N^\phi=-\chi\,,\qquad N^z=0\,,
\eeq
so that the metric in ZAMO-adapted form reads
\bea
ds^2&=&
-N^2 dt^2 +g_{ab}(dx^a-N^a dt)(dx^a-N^a dt)\nonumber\\ 
&=&-H\gamma^{2}dt^2 + \frac{r^2}{H\gamma^{2}}\left(d\phi-\chi dt \right)^2\nonumber\\
&+& e^{\Psi}\left(dr^2+dz^2\right), 
\eea
and naturally defines a ZAMO-adapted orthonormal (denoted by a hat) frame of 1-forms
\bea
\omega^{\hat 0}&=&\sqrt{H}\gamma dt\,,\nonumber\\
\omega^{\hat r}&=& e^{\Psi/2} dr\,,\nonumber\\   
\omega^{\hat \phi}&=& \frac{r}{\sqrt{H}\gamma}\left(d\phi-\chi dt \right)\,,\nonumber\\ 
\omega^{\hat z}&=&e^{\Psi/2}dz\,,
\eea
with dual frame
\bea
\label{ZAMO_frame}
e_{\hat 0}&=& n^\sharp \,,\qquad e_{\hat r}=e^{-\Psi/2}\partial_r\,,\nonumber\\ 
e_{\hat \phi}&=& \frac{\sqrt{H}\gamma }{r } \partial_\phi\,,\qquad e_{\hat z}=e^{-\Psi/2}\partial_z\,.
\eea
The fluid four velocity then reads
\bea
\label{Chi}
u=\gamma(n\pm v e_{\hat \phi}) \,,
\eea
so that $v=v(r,z)$ is the spatial three-velocity of the dust as measured by the ZAMOs.

%%%%%%

\section{Bonnor solution}

The case of a rigidly rotating dust is obtained by setting $H=1$, as discussed in Ref. \cite{Astesiano:2022gph}. For this subclass of solutions $\Omega$ is a constant, which can be always put consistently to zero by a proper rotation of the coordinate system.  Without any loss of generality we will then assume  $u=\partial_t$ for the fluid four-velocity. 
The request $\Omega=0$ imposes on $v(r,z)$ the constraint
\beq
\label{eq:constr_v_A}
v=-\frac{A}{r}\,,
\eeq
where $A=A(r,z)$ satisfies the equations
\bea
\label{GFequation1}
\frac{(\partial_{z}A)^2+(\partial_rA)^2}{r^2 e^{\Psi}} &=&8 \pi G\rho_m
\,, \\
\label{GFequation2}
\partial_{rr} A+ \partial_{zz} A- \frac{\partial_r A}{r}&=& 0 \,.
\eea
The function $\Psi$ satisfies the equations
\bea
\label{eqs_for_psi}
\Psi_{,r}= \frac{A^2_{,z}-A^2_{,r}}{2 r }\,,\qquad
\Psi_{,z}=-\frac{A_{,r} A_{,z}}{r }\,,
\eea
which have Eq. \eqref{GFequation2} as a compatibility condition.
Separable solutions to this equation (as functions of $r$ and $z$) will be reviewed in Appendix \ref{Solgenapp}.
However, Eq. \eqref{GFequation2} is a homogenous Grad-Shafranov equation \cite{Ruggiero:2023geh}, and admits separable solutions in the form of a multipole expansion by introducing polar coordinates on the  $r-z$ plane
\beq
\label{r_z_polar}
r=R\cos\alpha\,,\qquad z=R\sin\alpha\,.
\eeq
Such solutions write as
\beq
\psi(R,\alpha)=\sum_{n=2}^{\infty}\left(\alpha_{n}R^{n}+\beta_{n}R^{1-n} \right) \sin \alpha P^{1}_{n-1}(\cos \alpha) \label{eq:solsph2} 
\eeq
where $P^{1}_{n-1}(\cos \alpha)$ are the Legendre functions, and $\alpha_{n},\beta_{n}$ are arbitrary constants. We will consider below a particular example of rotating dust, namely Bonnor's solution \cite{Bonnor:1977}, corresponding to $\alpha_{2}=0$ and $\beta_{2}=m^{2}$. Accordingly, we have
\bea
A=\frac{m^2 \cos^2\alpha}{R}
\,,\qquad
\Psi= \frac{m^4 \cos^2\alpha\left(9\cos^2\alpha-8\right)}{8  R^4}\,.
\eea

\begin{figure}
\centering
\includegraphics[scale=0.3]{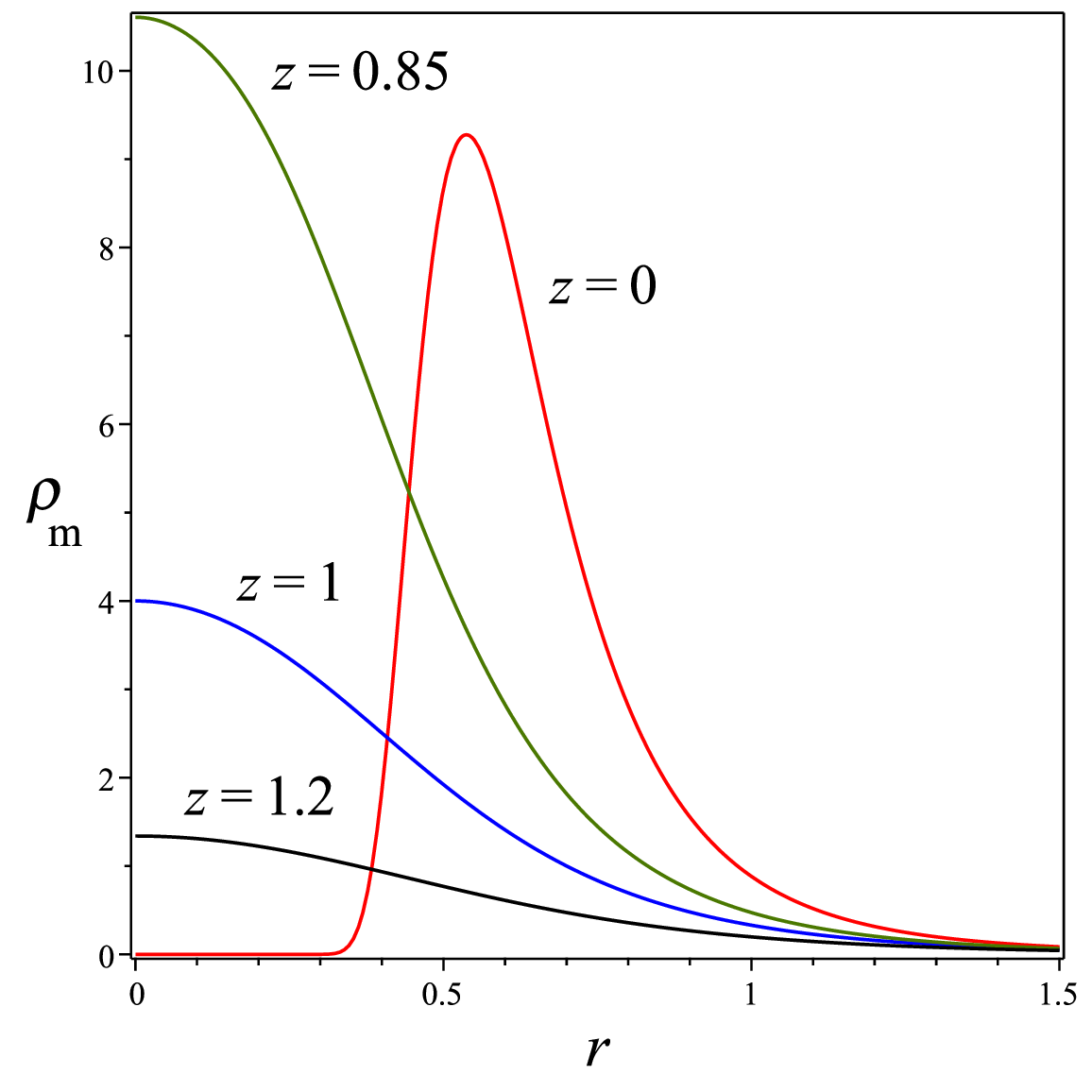}
\caption{
\label{fig:rhom} 
The behavior of the energy density $\rho_m(r,z)$ is plotted as a function of $r$ for fixed values of $z=[0,0.85,1,1.2]$, in units of $\kappa^{-1}$ and assuming $m=1$. 
}
\end{figure}

Bonnor's line element written in standard cylindrical-like coordinates is then given by
\bea
\label{MFD}
     ds^2&=& -dt^2-2 A  dt d\phi +\left(r^2- A^2\right)d\phi^2\nonumber\\
&+& e^{\Psi} \left(dr^2+dz^2\right)\,, 
\eea
with
\bea
\label{APsidefs}
A=\frac{m^2 r^2}{\left(r^2+z^2\right)^{3/2}}
\,,\qquad
\Psi= \frac{m^4 r^2\left(r^2-8 z^2\right)}{8  \left(r^2+z^2\right)^4}\,.
\eea
The associated dust energy density (see Fig. \ref{fig:rhom}) reads
\beq
\label{rho_m_eq}
   \kappa \rho_m = \frac{m^4 \left(r^2+4 z^2\right)}{ \left(r^2+z^2\right)^4}e^{-\Psi}\,,
\eeq
with $\kappa=8\pi G$ (actually $\kappa=8\pi$ after setting $G=1$ which is a standard choice here unless differently noted), entering the matter energy momentum tensor 
\beq
\label{T_munu}
T^{\mu\nu}=\rho_m u^\mu u^\nu = \rho_m \delta^\mu_0\delta^\nu_0\,.
\eeq
Here $m$ has the dimensions of a length, and it can be used to adimensionalize $t$,  $r$ and $z$,
\beq
\label{resc_vars}
\hat t=\frac{t}{m}\,,\qquad \hat r =\frac{r}{m}\,,\qquad \hat z=\frac{z}{m}\,.
\eeq
Sometimes we will find convenient to use the polar coordinates \eqref{r_z_polar}, with $R$ adimensionalized as $\hat R=R/m$.
$m$ can also be used to rescale and adimensionalize all metric quantities (and associated tensors). For example, if $X(m,r,z)$ has the dimensions of $(length)^n$ then
$X(m,r,z)=m^n \hat X(\hat r,\hat z)$, with $\hat X$ dimensionless.

The congruence of world lines tangent to the fluid four velocity  $u=\partial_t$ is geodesic (vanishing acceleration $a(u)=0$) and shear-free (vanishing expansion $\theta(u)=0$), with nonvanishing vorticity vector (see, e.g., Ref. \cite{Jantzen:1992rg} for definitions and notations of kinematical quantities)
\beq
\omega(u)=\frac{e^{-\Psi}m^2}{(r^2+z^2)^{5/2}}\left[3 rz\partial_r-(r^2-2z^2)\partial_z\right]\,.
\eeq
The latter turns out to be aligned with the $z$ axis when evaluated on the symmetry plane ($z=0$),
\beq
\omega(u)|_{z=0}=-\frac{m^2}{r^3}\, e^{-\frac{m^4}{8 r^4 }}\partial_z \,,
\eeq
and exponentially decaying asymptotically for large $r$.
Moreover, 
\beq
|\omega(u)|^2=\kappa \rho_m\,,
\eeq
as noticed in Ref. \cite{Bonnor:1977}.

Unfortunately, Bonnor's spacetime admits a pathological domain. In fact, the vanishing of the $g_{\phi\phi}$ metric component is responsible for the appearance of an unphysical region characterized by the existence of CTCs
\beq
r^2- A^2=0\,,
\eeq
leading to the following toroidal  boundary 
\beq
\label{zctc}
z_{\rm ctc}=\pm\sqrt{m^{4/3}r^{2/3}-r^2}\,,
\eeq
or, equivalently, in terms of the rescaled and dimensionless variables \eqref{resc_vars}
\beq
\label{hat_zctc}
\hat z_{\rm ctc}=\pm\sqrt{\hat r^{2/3}-\hat r^2}\,.
\eeq
On the symmetry plane $\hat z_{\rm ctc}=0$, implying that the toroidal region reduces to the circle $\hat r_{\rm ctc}=1$.
Even if this can be interesting as a matter of principle, in the following discussion we will not consider physical phenomena in this region.

\subsection{Curvature invariants}

The Kretschmann invariant for this metric is given by
\bea
K= R^{\alpha\beta\gamma\delta}R_{\alpha\beta\gamma\delta}
= {\mathcal K}\frac{m^4 e^{-2\Psi}}{(r^2+z^2)^{12}}\,,
\eea
with
\bea
{\mathcal K}&=&-36r^{16}-360r^{14} z^2+(8 m^4-1512 z^4)r^{12}\nonumber\\
&-&3 z^2 (1176 z^4+m^4)r^{10}\nonumber\\
&+& (-141 m^4 z^4-\frac14 m^8-5040 z^8)r^8\nonumber\\
&-&z^2 (4536 z^8+293 m^4 z^4+3 m^8)r^6\nonumber\\
&-&3 z^4 (840 z^8+61 m^4 z^4+4 m^8)r^4\nonumber\\
&-&4 z^6 (198 z^8-3 m^4 z^4+4 m^8)r^2\nonumber\\
&+& 4 z^{12} (-27 z^4+8 m^4)
\,,
\eea
and $m^4K$ dimensionless.
Taking the limit $r\to \infty$ at fixed $z=$constant yields
\beq
K= -\frac{36 m^4}{r^8}+\frac{72 z^2 m^4}{r^{10}}+O\left(\frac{1}{r^{12}}\right)\,,
\eeq
whereas for $r\to0$ 
\bea\fl\quad
K&=&\frac{4m^4(8m^4-27z^4)}{z^{12}}\qquad\nonumber\\
&+&
\frac{12m^4(4m^8-49z^4m^4+42z^8)}{z^{18}}r^2+O(r^4)
\,.\qquad
\eea

The second quadratic invariant is given by
\bea
{}^*K= R_{\alpha\beta\gamma\delta}{}^*R^{\alpha\beta\gamma\delta}
= {}^*{\mathcal K}\frac{m^6 e^{-2\Psi}}{(r^2+z^2)^{21/2}}\,z\,,
\eea
with
\bea
{}^*{\mathcal K}&=&-18 r^{10}-120z^2 r^8 -\left(\frac32 m^4+300 z^4\right)r^6
\nonumber\\
&-&\left(12 m^4 +360 z^4\right)z^2 r^4\nonumber\\
&-& (24 m^4 +210 z^4)z^4r^2-48 z^{10}
\,,
\eea
and $m^4{}^*K$ dimensionless.

For large values of the radial coordinate at fixed $z$ it behaves as
\beq
{}^*K= -\frac{18 z m^6}{r^{11}}+\frac{69 z^3 m^6}{r^{13}}+O\left(\frac{1}{r^{15}}\right)\,,
\eeq
whereas for $r\to0$
\beq
{}^*K=-\frac{48m^6}{z^{10}}-\frac{6m^6(-49z^4+20m^4)}{z^{16}}r^2+O(r^4)\,.\qquad
\eeq

Therefore, both invariants diverge approaching $r=0$ for $z\to0$ only, implying that the singularity at the origin has a directional character, as already discussed by Bonnor himself.
In fact, when approaching it along the lines $z=kr$, $r=$constant the argument $\Psi$ of the exponential term changes sign for $1-8k^2=0$, so that the curvature invariants as well as the density exhibit different behavior for $\sqrt{r^2+z^2}\to0$.
No other singularities are present, so that the Bonnor spacetime represents a rotating dust cloud containing an isolated singularity. 
The latter is not generated by a matter distribution at the center, but rather by the vorticity of the fluid which generally has a divergent behavior there.
For instance, for $r\to0$ at fixed $z$
\beq
|\omega(u)|^2=\kappa \rho_m
=\frac{4m^4}{z^6}+\frac{m^4(-15z^4+4m^4)}{z^{12}}r^2+O(r^4)\,,
\eeq
which diverges for $z\to0$.

We will come back on these invariants after having discussed about timelike geodesics in subsection 3.3.

\subsection{Spectral Petrov type properties}

Using $e_{\hat 0}=n^\sharp$ with the corresponding natural adapted spatial frame one can form the null Newman-Penrose tetrad
\bea\fl\quad
l_{\rm NP}&=&\frac{1}{\sqrt{2}}(e_{\hat 0}+e_{\hat \phi})\,,\qquad n_{\rm NP}=\frac{1}{\sqrt{2}}(e_{\hat 0}-e_{\hat \phi})\,,\qquad\nonumber\\
m_{\rm NP}&=&\frac{1}{\sqrt{2}}(e_{\hat r}+i e_{\hat z})\,.
\eea
The associated Weyl scalars are $\psi_1=0=\psi_3$, and
\beq
m^2e^{ \Psi} \psi_2=  \frac{1}{4\hat R^4}\left(- 3i \sin \alpha   +\frac{ 5-3\cos(2\alpha)}{12 \hat R^2}\right)  \,, 
\eeq
while the expressions for $\psi_0$ and $\psi_4$ are of the type
\bea
\label{psi0_and_psi4}
m^2e^{ \Psi}\psi_0&=& \frac{C^0_6 \hat R^6+C^0_4 \hat R^4+C^0_2 \hat R^2 +C^0_0}{\hat R^8 (\cos \alpha -\hat R^2)}\,,\nonumber\\
m^2e^{ \Psi}\psi_4&=& \frac{C^4_6 \hat R^6+C^4_4 \hat R^4+C^4_2 \hat R^2 +C^4_0}{\hat R^8 (\cos \alpha +\hat R^2)}\,,
\eea
where the various coefficients $C^0_n$ and $C^4_n$ are listed in Table \ref{tab:table1}.
The metric is thus of Petrov type I.

\begin{table}  
\caption{\label{tab:table1} List of the  various coefficients $C^0_n$ and $C^4_n$ entering Eqs. \eqref{psi0_and_psi4}.
}
\begin{ruledtabular}
\begin{tabular}{l|l}
$ C^0_6$ & $96\cos(\alpha)+96i\sin(\alpha)-480i\sin(3\alpha)-480\cos(3\alpha) $\\
$ C^0_4$ & $ -144\cos(2\alpha)+40-312\cos(4\alpha)-144i\sin(2\alpha)$\\
&$-312i\sin(4\alpha)$\\
$ C^0_2$ & $ 42i\sin(\alpha)-48 \cos(3\alpha)-36i\sin(5\alpha)-48i\sin(3\alpha)$\\
&$+54i\sin(7\alpha)+54\cos(7\alpha)+30\cos(\alpha)-36\cos(5\alpha)$\\
$ C^0_0$ & $7-18\cos(4\alpha)-11\cos(2\alpha)-9i\sin(2\alpha)+27 \cos(6\alpha)$\\
&$+27i\sin(8\alpha)+27i\sin(6\alpha)+27\cos(8\alpha)-18i\sin(4\alpha) $\\
\hline
$ C^4_6$ & $-480\cos(3\alpha)+96\cos(\alpha)+480i\sin(3\alpha)-96i\sin(\alpha) $\\
$ C^4_4$ & $ 312\cos(4\alpha)-312i\sin(4\alpha)+144\cos(2\alpha)$\\
&$ -144i\sin(2\alpha)-40$\\
$ C^4_2$ & $-42i\sin(\alpha)+48i\sin(3\alpha)+54\cos(7\alpha)+36i\sin(5\alpha)$\\
&$-54i\sin(7\alpha)+30\cos(\alpha)-36\cos(5\alpha)-48\cos(3\alpha) $\\
$ C^4_0$ & $-7-9i\sin(2\alpha)+27i\sin(8\alpha)-18i\sin(4\alpha)+27i\sin(6\alpha)$\\
&$+18\cos(4\alpha)+11\cos(2\alpha)-27\cos(8\alpha)-27\cos(6\alpha) $\\
\end{tabular}
\end{ruledtabular}
\end{table}

\subsection{Timelike geodesics}

\begin{figure}
\centering
\includegraphics[scale=.3]{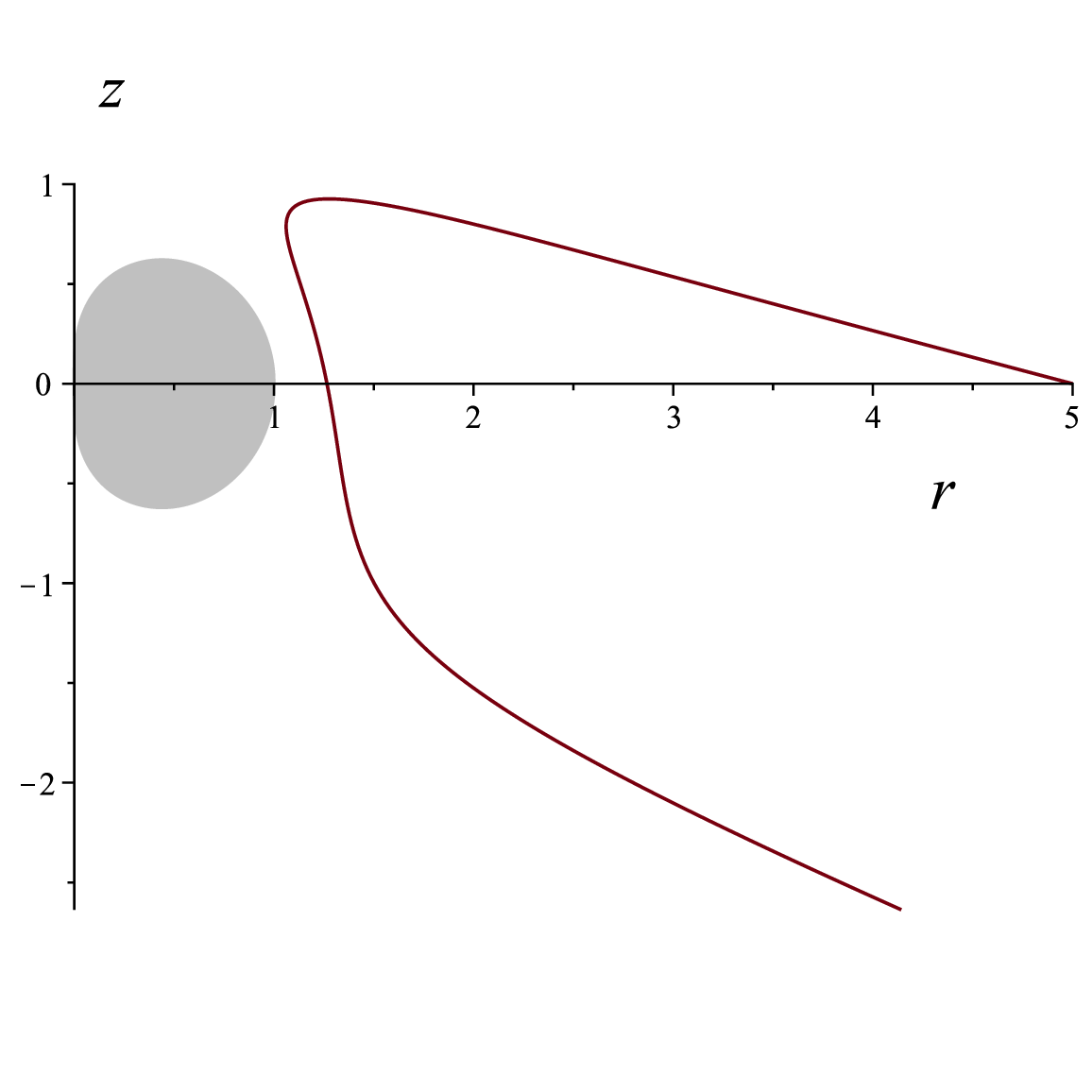}
\caption{
\label{fig:plot_noneq1} 
Example of numerical integration of the orbit, Eq. \eqref{general_geos}, projected on the $r$-$z$ plane for the following choice of parameters: $m=1$, $E=1.5$, and $L=-1.8$.
Initial conditions are  chosen so that $R(0) = 5$, $\alpha(0) = 0$, $\phi(0) = 0$, $t(0) = 0$, $\frac{d\alpha}{d\tau}\big\vert_{\tau=0} = 0.055$ and $\frac{dR}{d\tau}\big\vert_{\tau=0} \approx -1.0412$, the latter coming from the normalization condition for timelike orbits. The shaded region (grey online) denotes the unphysical region with boundary \eqref{zctc} where CTCs exist.  
}
\end{figure}

Let us consider the (timelike, geodesic) motion of a massive particle with four-velocity  $U=U^\alpha \partial_\alpha$. 
The conserved quantities $E=-U_t$ (energy per unit of particle mass) and $L=U_\phi$ (angular momentum per unit of particle mass) are related to the contravariant components of the four velocity as
\beq
\label{EL_defs}
    E=U^t +A U^\phi\,,\qquad L=A U^t+ (r^2-A^2)  U^\phi\,,
\eeq
so that
\beq
    U^t=E -A \frac{L+E A}{r^2}\,, \qquad  U^\phi= \frac{L+E A}{r^2}\,,
\eeq
with $A$ given in Eq. \eqref{APsidefs}.
Note that $U^\phi$ vanishes when $A=-L/E$ (even for $r=$ constant motion).  
The normalization condition $U\cdot U=-1$ implies
\beq
\label{normaliz}
g_{zz}  (U^z)^2+ g_{rr} (U^r)^2=E^2-1-\frac{(L+E A)^2}{r^2}\,. 
\eeq
\begin{widetext}
The geodesic equations written in $(t,R,\phi,\alpha)$ coordinates, Eq. \eqref{r_z_polar},  then read
\bea
\label{general_geos}\fl
\frac{dt}{d\tau}&=& \frac{ -\cos^2\alpha   m^4 E-L R m^2+R^4 E }{R^4}
\,,\nonumber\\
\fl
\frac{d\phi}{d\tau}&=& \frac{ \cos^2\alpha m^2 E+R  L }{R^3\cos^2\alpha } 
\,,\nonumber\\
\fl
\frac{d^2R}{d\tau^2}&=& \frac{m^4\cos^2\alpha (9\cos^2\alpha-8)}{4R^5} \left(\frac{dR}{d\tau}\right)^2
+\frac{m^4\sin\alpha \cos\alpha (-4+9\cos^2\alpha)}{2R^4} \frac{dR}{d\tau}\frac{d\alpha}{d\tau}\nonumber\\
\fl
&-&\frac{(-8 m^4\cos^2\alpha  -4R^4+9 m^4\cos^4\alpha  )}{4R^3}  \left(\frac{d\alpha}{d\tau}\right)^2\nonumber\\
\fl
&+&\frac{(2 m^4 E^2\cos^4\alpha+R^2 L^2+3R E L m^2\cos^2\alpha)} {R^5\cos^2\alpha} e^{\frac{ - m^4\cos^2\alpha (9\cos^2\alpha -8)}{8 R^4}}
\,,\nonumber\\
\fl
\frac{d^2\alpha}{d\tau^2}&=& -\frac{ m^4\sin\alpha \cos\alpha (-4+9\cos^2\alpha)}{4R^6}\left[\left(\frac{dR}{d\tau}\right)^2-R^2
\left(\frac{d\alpha}{d\tau}\right)^2\right]\nonumber\\
\fl
&-&\frac{(8 m^4\cos^2\alpha -9 m^4\cos^4\alpha  +4R^4)}{2R^5}  \frac{dR}{d\tau}\frac{d\alpha}{d\tau}\nonumber\\
\fl
&+&
\frac{\sin\alpha  (-R^2 L^2+m^4 E^2\cos^4\alpha )}{\cos^3 \alpha  R^6}e^{\frac{ - m^4\cos^2\alpha (9\cos^2\alpha -8)}{8 R^4}} \,, 
\eea
\end{widetext}
and can be integrated numerically.
As an example, we have shown in Fig. \ref{fig:plot_noneq1} the behavior of the integrated orbit in the $r$-$z$ section. 
The figure also displays the location of the unphysical region containing CTCs.
It is interesting to note how the orbit is somehow \lq\lq repelled" by this region as a typical feature.
However, there exist conditions on the energy and angular momentum parameters such that particles can enter the CTC region, as discussed below.

\subsubsection{Evaluating curvature invariants along geodesics}

\begin{figure} 
\includegraphics[scale=.25]{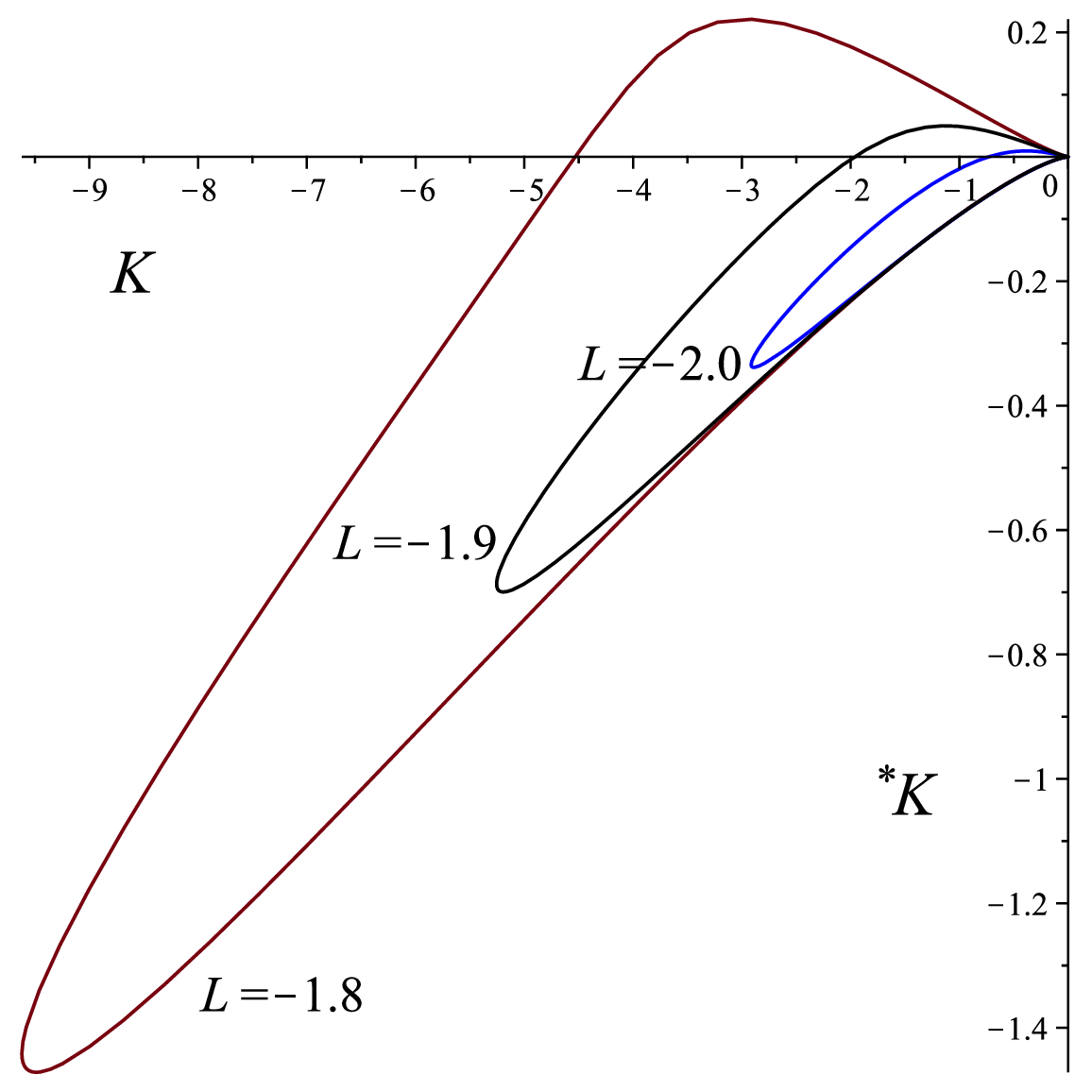} 
\caption{
\label{fig:7}
The curvature invariants  ${}^*K$ vs $K$  evaluated along three non-equatorial geodesics: the one of Fig. \ref{fig:plot_noneq1} (with $L=-1.8$), and two similar hyperboliclike orbits corresponding to the same set of parameters as well as initial conditions, except for slightly different values of the angular momentum, $L=-1.9$ and $L=-2.0$. 
The initial conditions are such that the geodesics start on the symmetry plane where ${}^*K=0$ at a given radius with $K<0$, and approach spatial infinity, where both invariants vanish, so that the curves all terminate at the origin.    
}
\end{figure}

We can evaluate the curvature invariants $K$ and ${}^*K$, previously given as functions of the coordinates $r$ and $z$, along the geodesic motion discussed above so that they actually become functions of the orbital parameters, including proper time.
A parametric plot ${}^*K$ vs $K$  eliminates the dependence on proper time, and gives a (gauge-invariant) curve, or better families of curves, only depending on $E$, $L$ and $m$.
This is an interesting way to look at the geometrical properties of the spacetime too.
Fig \ref{fig:7} shows a geometrical information concerning curvature invariants associated with the geodesic of Fig. \ref{fig:plot_noneq1} as an example. From the plots one sees that the curves ${}^*K$ vs $K$ intersect the horizontal axis (${}^*K=0$) as soon as the geodesic crosses the symmetry plane $z=0$. Furthermore, each curve terminates at the origin, since both invariants vanish at spatial infinity.

\subsubsection{Motion on the symmetry plane}

To study the motion on the symmetry plane ($z=0$, $U^z=0$), where $r_{\rm ctc}=m$, it is convenient to introduce the associated effective potential for radial motion
\beq
 g_{rr}(U^r)^2=\frac{ g_{\phi\phi}}{r^2}(E-V_{\rm eff}^{r\,+})(E-V_{\rm eff}^{r\,-})\,,
\eeq
where 
\beq
V_{\rm eff}^{r\,\pm}=\frac{rm^2 L}{r^4-m^4}\pm\frac{r^2\sqrt{r^4+L^2r^2-m^4}}{r^4-m^4}\,,
\eeq
which tends to the limiting value $\pm1$ for large values of $r$. 
Consider, for instance, the positive branch of the effective potential. 
Taking the limit $r\to m$, i.e., as the location of the CTC boundary is approached, the latter behaves as
\bea
V_{\rm eff}^{r\,+}&=&\frac14|L|(1+{\rm sign}(L))(r-m)^{-1}\nonumber\\
&+&\frac12\left[\frac{m}{|L|}+\frac{|L|}{4m}(3-{\rm sign}(L))\right]
+O(r-m)\,,\nonumber\\
\eea
so that the CTC boundary is a potential barrier for all particles with positive values of the angular momentum.

Imposing the condition $U^r=0$ in Eq. \eqref{normaliz} gives the turning points of the radial motion. Solving for $r$ yields four roots $r^*_i$ ($i=1,\dots,4$) 
\beq
\label{tprad}
r^*_i=\frac{\sigma L}{2\sqrt{E^2-1}}+\frac{\epsilon}{2(E^2-1)^{1/4}}\sqrt{\frac{L^2}{\sqrt{E^2-1}}+4m^2\sigma E}\,,
\eeq
where $\sigma=\pm$, $\epsilon=\pm$ are independent signs, and $E>1$.
The number of turning points (i.e., positive real roots, for fixed values of the energy) determines the particular radial motion, either bound or hyperboliclike.
It is convenient to treat the cases $L>0$ and $L<0$ separately.
By using the rescaled variables
\beq
\hat r_i^*=\frac{r^*_i}{m}\,,\qquad  \hat L=\frac{L}{m}\,,
\eeq
as well as the notation
\beq
\xi=\frac{\hat L}{2|\sinh\beta|}\,,\qquad  E=\cosh \beta\,,
\eeq
Eq. \eqref{tprad} can be conveniently rewritten as
\beq
\label{tpradnew}
\hat r^*_i\equiv {\mathcal R}(\sigma,\epsilon)=\sigma \xi+\epsilon\sqrt{\xi^2+\sigma|\coth \beta|}\,,
\eeq
with ${\rm sign}(\xi)={\rm sign}(L)$.

For positive values of the angular momentum there are four real roots, provided that the argument of the square root in the second term of Eq. \eqref{tpradnew} is positive, i.e., for $\xi>\sqrt{|\coth \beta|}$, but only one is positive, namely ${\mathcal R}(1,1)$.
For $\xi<\sqrt{|\coth \beta|}$, instead, there are two complex conjugate roots, and two real roots, one of which is negative, ${\mathcal R}(1,-1)$, and the other positive, ${\mathcal R}(1,1)$.
Hence, for $L>0$ there is always a single turning point at $\hat r_{\rm tp}={\mathcal R}(1,1)$.
The motion can be only hyperboliclike in this case.

The case $L<0$ is more interesting.
In fact, for $|\xi|>\sqrt{|\coth \beta|}$ there are four real roots, one of them is negative, implying that there are at most three tuning points, depending on the chosen value of the energy, i.e., $\hat r_{1\,\rm tp}={\mathcal R}(1,1)$, $\hat r_{2\,\rm tp}={\mathcal R}(-1,-1)$, and $\hat r_{3\,\rm tp}={\mathcal R}(-1,1)$. 
In contrast, for $|\xi|<\sqrt{|\coth \beta|}$ there are two real roots (one positive, ${\mathcal R}(1,1)$, the other negative, ${\mathcal R}(1,-1)$) and two complex conjugate roots, so that the only turning point is $\hat r_{\rm tp}={\mathcal R}(1,1)$.
Therefore, for positive values of the angular momentum bound motion is also allowed.

\begin{figure}\[
\begin{array}{cc}
\includegraphics[scale=.20]{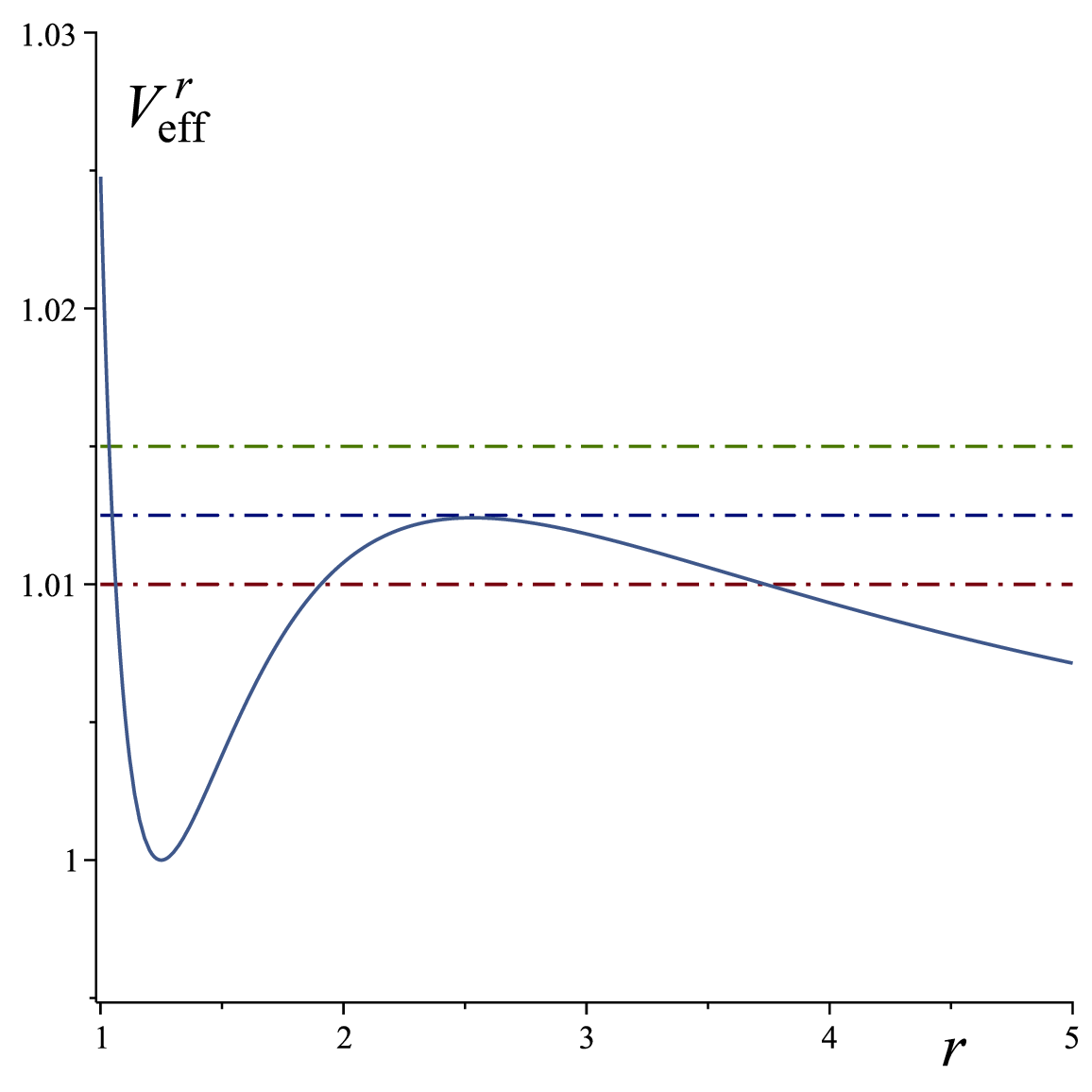}&
\includegraphics[scale=.20]{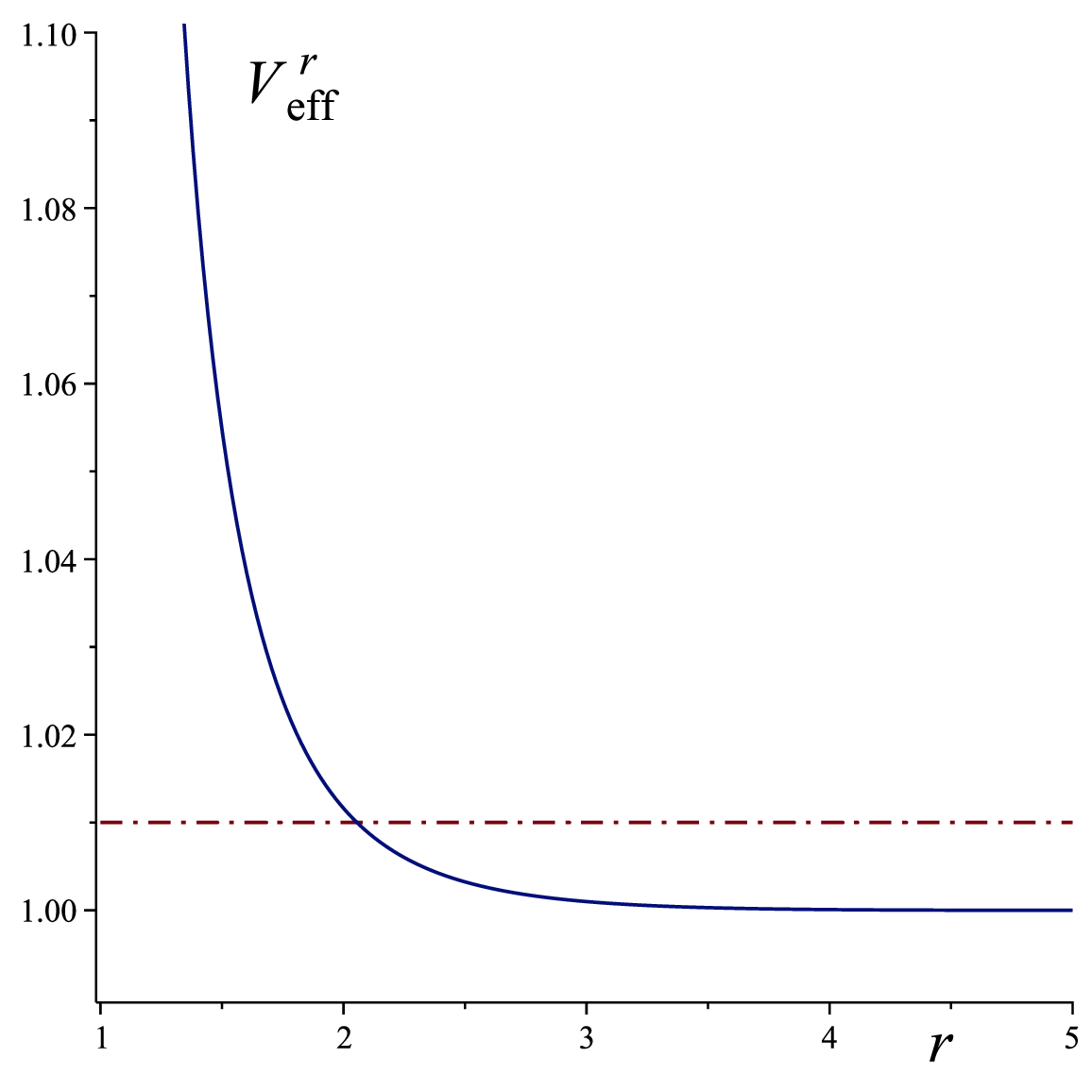}\cr
(a) & (b) \cr
\end{array}
\]
\caption{
\label{fig:Veff}
Typical behavior of the (positive branch of the) effective potential for radial motion as a function of the radial coordinate for $m=1$ (implying that the CTC boundary is at $r_{\rm ctc}=1$) and different values of the particle's angular momentum parameter.
The horizontal lines correspond to different values of the particle's energy parameter, and intersect the potential curve at the turning points.
Left panel: $L = -0.8$; $E=1.01$ (three distinct roots), $E\approx1.0124$ (three roots, two of them coinciding), and $E=1.015$ (one root).
Right panel: $L = -0.2$; $E=1.01$ (one root).
}
\end{figure}

The typical behavior of the (positive branch of the) effective potential as a function of the radial coordinate $r$ is shown in Fig. \ref{fig:Veff} for selected (negative) values of the particle's angular momentum parameter.
Horizontal lines corresponding to fixed values of the particle's energy parameter intersect the potential curve at the turning points of the radial motion. 

\begin{figure}\[
\begin{array}{cc}
\includegraphics[scale=.20]{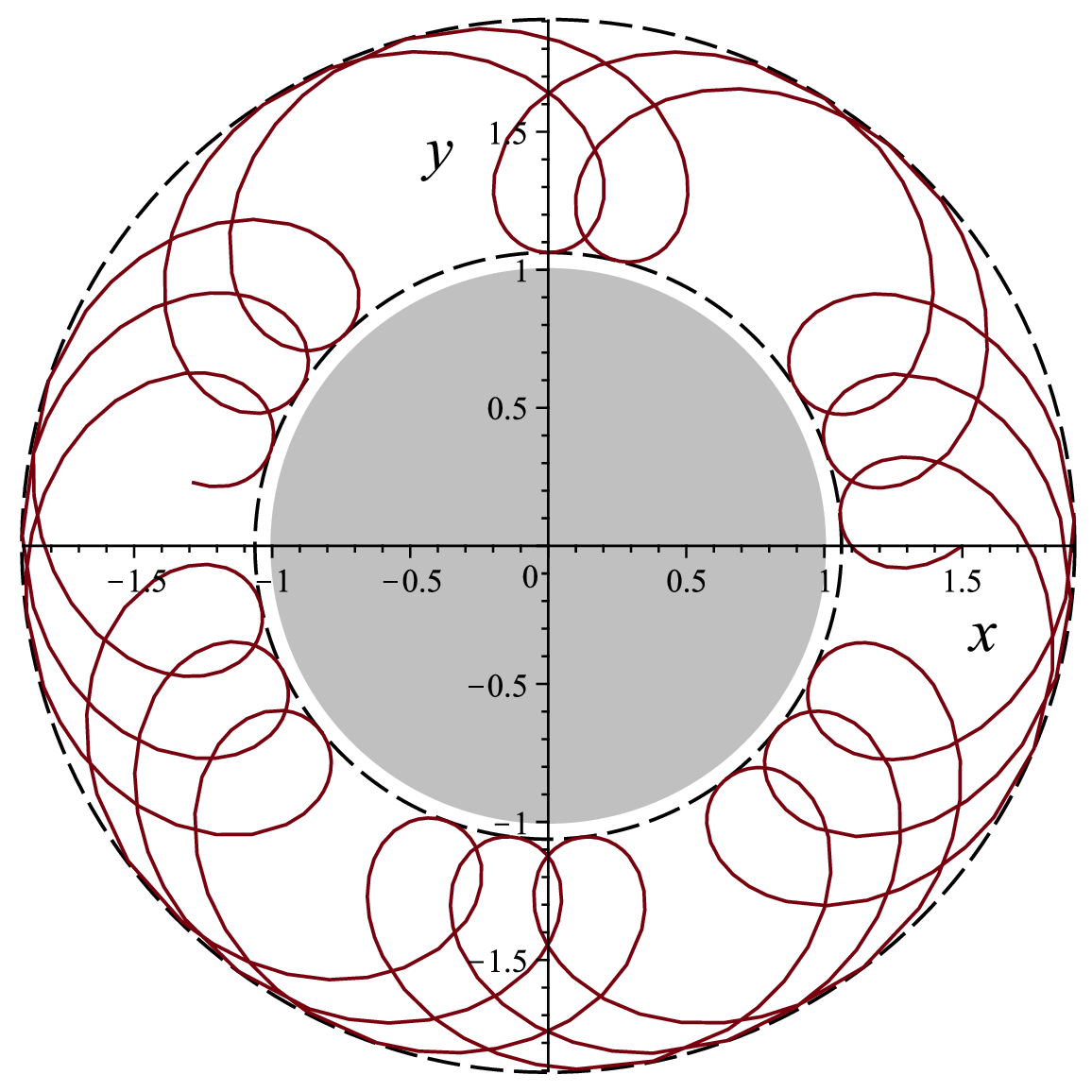}&
\includegraphics[scale=.20]{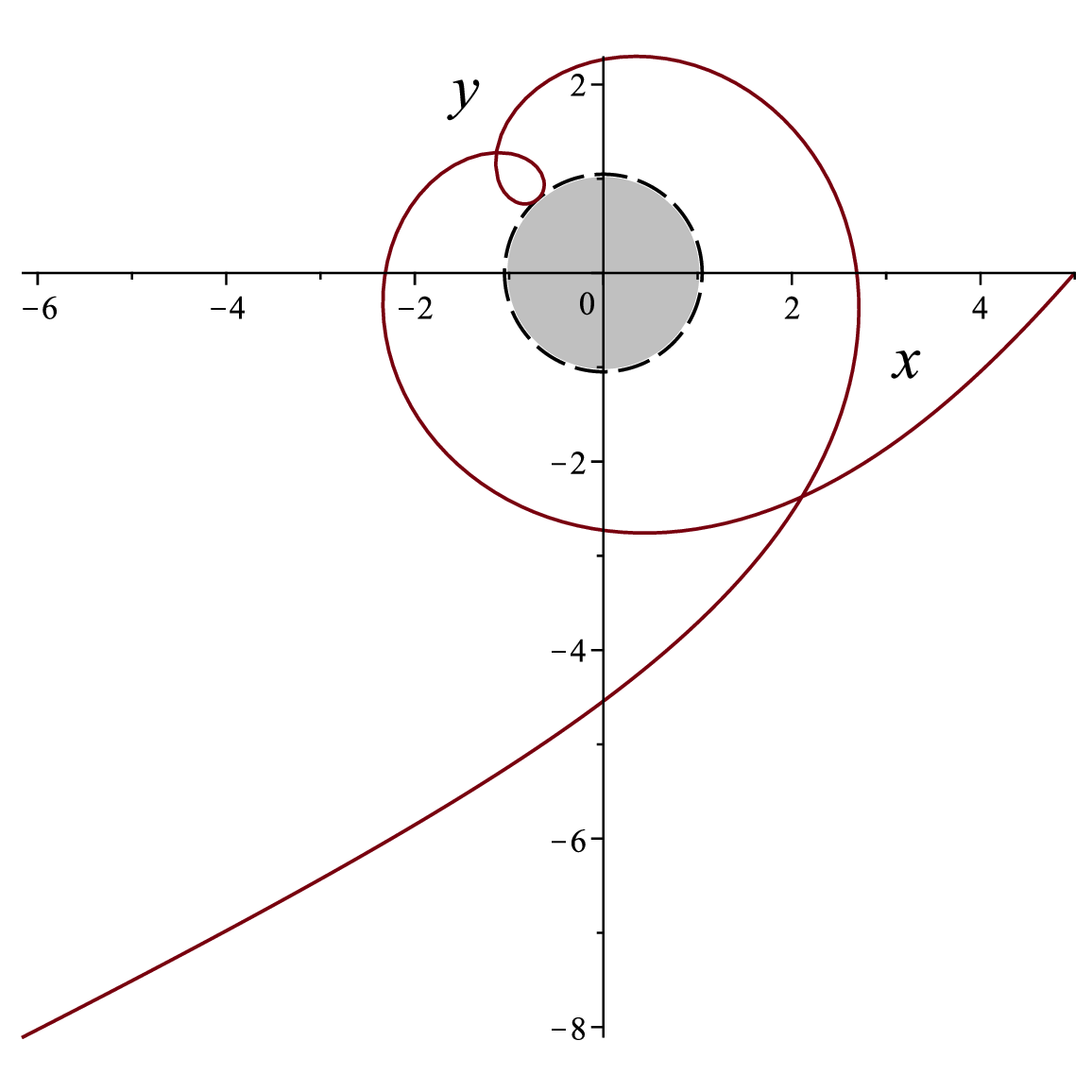}\cr
(a) & (b) \cr
\includegraphics[scale=.20]{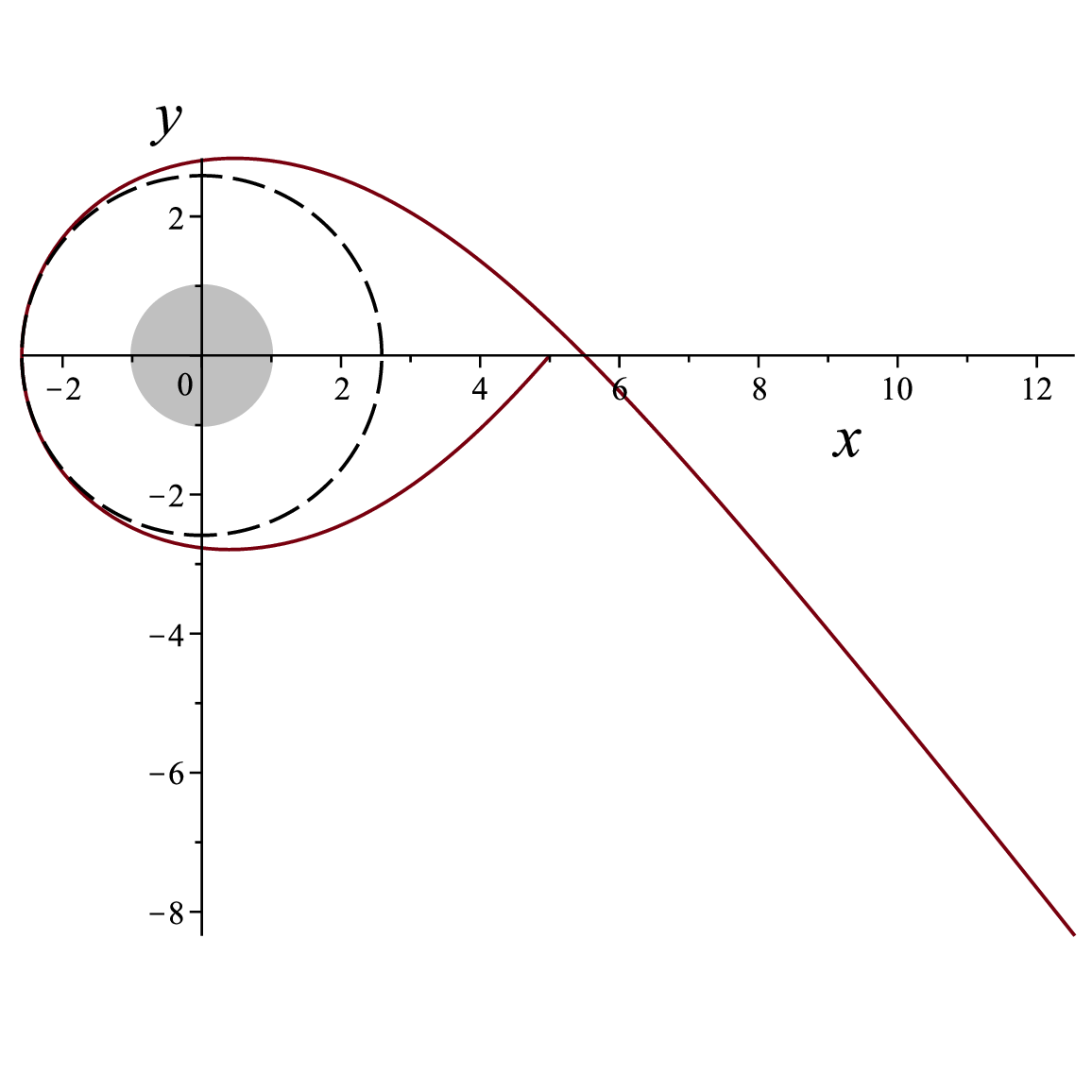}&
\includegraphics[scale=.20]{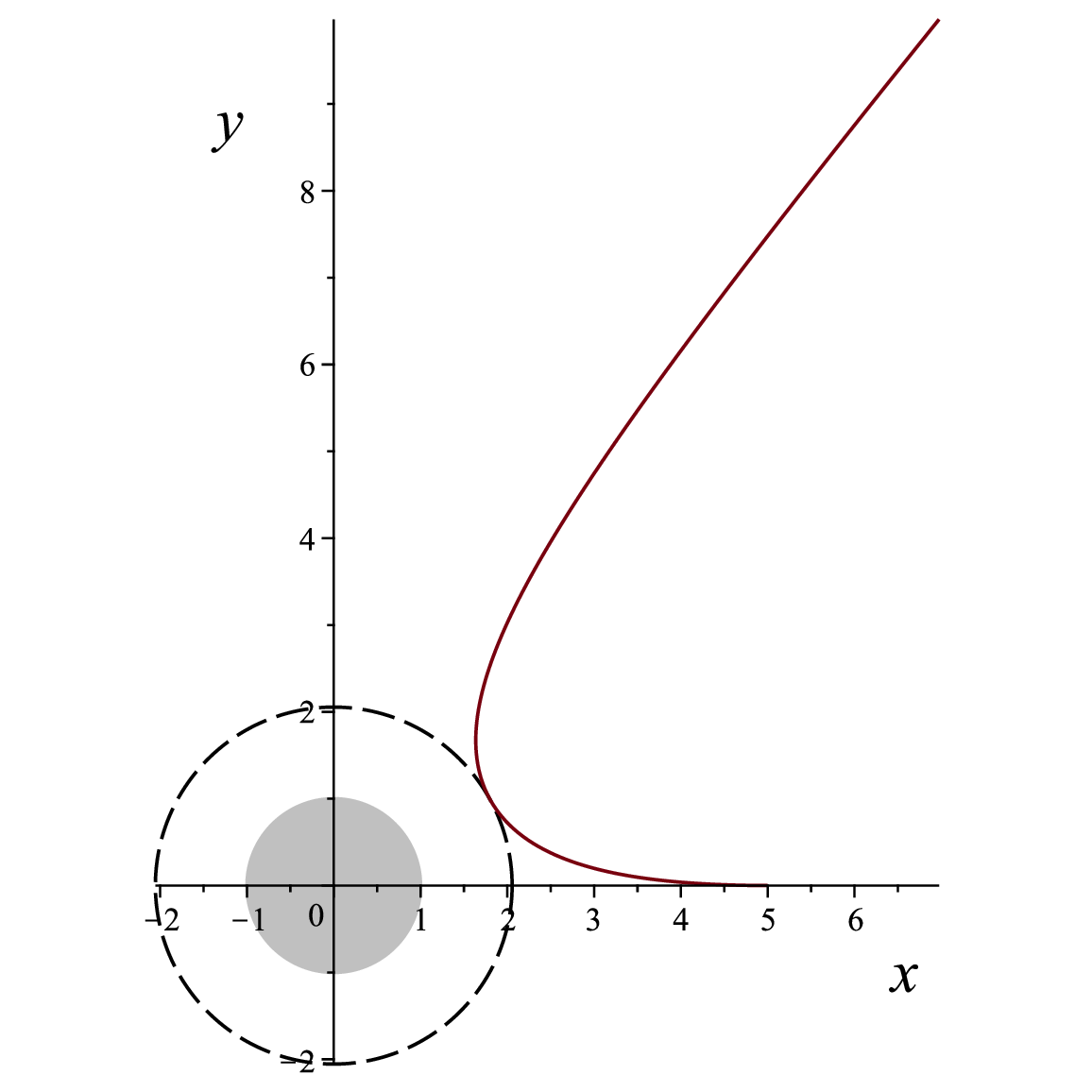}\cr
(c) & (d) \cr
\end{array}
\]
\caption{
\label{fig:equat}
Examples of numerical integration of the geodesic orbits in the symmetry plane $z=0$.
In all plots we have set the value of the parameter $m=1$, and the initial condition for $t$, $\phi$ and $\alpha$ to the values $t(0) = 0$, $\phi(0) = 0$, $\alpha(0) = 0$. The shaded region (grey online) containing CTCs is always displayed, and never crossed by the orbits.
Panel (a): parameter choice: $E=1.01$, $L=-0.8$; initial conditions: $R(0) = 1.5$,  $\frac{dR}{d\tau}\big\vert_{\tau=0} \approx -0.1125$, $\frac{d\alpha}{d\tau}\big\vert_{\tau=0} = 0$; turning points: $r_1^*\approx1.0625$, $r_2^*\approx1.9069$.
Panel (b): parameter choice: $E = 1.0125$, $L = -0.8$; initial conditions: $R(0) = 5$,  $\frac{dR}{d\tau}\big\vert_{\tau=0} \approx -0.1043$, $\frac{d\alpha}{d\tau}\big\vert_{\tau=0} = 0$; turning points: $r^*\approx1.0479$ (other roots corresponding to imaginary values).
Panel (c): parameter choice: $E = 1.0124$, $L = -0.8$; initial conditions: $R(0) = 5$,  $\frac{dR}{d\tau}\big\vert_{\tau=0} \approx -0.1033$, $\frac{d\alpha}{d\tau}\big\vert_{\tau=0} = 0 $; turning points: $r^*\approx2.5865$ (the one corresponding to the largest values of $r$).
Panel (d): parameter choice: $E = 1.01$, $L = -0.2$; initial conditions: $R(0) = 5$,  $\frac{dR}{d\tau}\big\vert_{\tau=0} \approx -0.1418$, $\frac{d\alpha}{d\tau}\big\vert_{\tau=0} = 0$; turning points: $r^*\approx2.0554$ (other roots corresponding to imaginary values).
}
\end{figure}

Fig. \ref{fig:equat} shows some examples of incoming orbits from a given initial value for the radial distance and the same choice of parameters as in Fig. \ref{fig:Veff}.

Let us conclude this section by discussing the issue of CTC avoidance in the case $L<0$.
The particle will enter the CTC region during its motion if the innermost radial turning point is less than the location of the CTC boundary, i.e., $\hat r_{1\,\rm tp}={\mathcal R}(1,1)\leq 1$.
The condition for CTC avoidance is then 
\beq
-|\xi|+\sqrt{\xi^2+|\coth \beta|}>1\,,
\eeq
so that
\beq
|\coth \beta|>1+2|\xi|\,.
\eeq
Turning to the original variables the latter condition writes as
\beq
|\hat L|< E-\sqrt{E^2-1}\,.
\eeq

\subsubsection{Bound motion}

Bound motion is conveniently studied by using an eccentricity-semi-latus rectum parametrization, which is familiar from the Newtonian description of the binary dynamics
\beq
r=\frac{p}{1+\epsilon\cos\chi}\,,
\eeq
where $p>m$ and $0\leq \epsilon<1$ are the semilatus rectum and eccentricity, respectively, and $\chi$ the polar angle in the plane of the orbit.
The motion is confined between a minimum value $r_{\rm peri}=p/(1+\epsilon)$ for $\chi=0$ and a maximum value $r_{\rm apo}=p/(1-\epsilon)$ for $\chi=\pi$ of the radial coordinate.
From Eq. \eqref{tprad} one can then express the particle energy and angular momentum parameters in terms of $(p,\epsilon)$ as follows
\bea
E&=&\frac{p^2}{\sqrt{p^4-m^4\epsilon^2(1-\epsilon^2)^2}}
\,,\nonumber\\
L&=&-\frac{m^2(1+\epsilon^2)}{p}E
\,.
\eea
The equation for the radial variable $r$ is thus replaced by the following equation for the angular variable $\chi$
\bea\fl\qquad
\frac{d\chi}{d\tau}&=&\frac{m^4E^2}{p^5}e^{-\frac{m^4(1+\epsilon\cos\chi)^4}{8p^4}}\epsilon\sin\chi(1+\epsilon\cos\chi)^2\times \nonumber\\
&&\left[\left(1-\epsilon^2+\epsilon\cos\chi\right)^2-\epsilon^2\right]\,,
\eea
which can be used to express $t$ and $\phi$ as functions of $\chi$.
One can then compute the period $T_r$ of the radial motion and the corresponding full variation $\Phi$ of the azimuthal angle 
\bea
T_r&=&\oint dt=2\int_0^{\pi}\frac{dt}{d\chi}d\chi
\,,\nonumber\\
\Phi&=&\oint d\phi=2\int_0^{\pi}\frac{d\phi}{d\chi}d\chi\,.
\eea
The associated radial and azimuthal frequencies are given by $\Omega_r=2\pi/T_r$ and $\Omega_{\phi}=\Phi/T_r$, respectively.
Analytical expressions for the above quantities can be obtained in the limit of small eccentricity ($\epsilon\ll1$) only as power
series expansions. The leading order terms turn out to be 
\bea
\Omega_r&=&e^{-\frac{m^4}{8p^4}}\frac{m^4}{2p^5\ln(2)}\pi \epsilon
+O(\epsilon^3)
\,,\nonumber\\
\Omega_{\phi}&=&\frac{m^2}{p^3\ln(2)}\left[2-3\ln(2)-(1-\ln(2))\frac{m^4}{2p^4}\right]\epsilon^2\nonumber\\
&+& O(\epsilon^4)\,.
\eea
The bound orbit of Fig. \ref{fig:equat} (a) corresponds to the parameter choice $p\approx1.3646$ and $\epsilon\approx0.2844$. 
The particle bounces many times between the pericenter and the apocenter before completing a full revolution around the origin.
It is interesting to note that the orbit does not reduce to a circular orbit in the limit $\epsilon\to0$.
In fact, in that limit $r$ goes to a constant value, with $E\to1$ and $L\to-{m^2}/{p}$, whereas the frequencies $\Omega_r$ and $\Omega_\phi$ both vanish, implying that eventually the particle comes to rest.\\

Finally, circular orbits are obtained by imposing the condition $\partial_rV_{\rm eff}^{r\,\pm}=0$, leading to analytical expression for the corresponding radii
\beq
\label{rcircsols}
r_{1\,\rm circ}=-\frac{m^2}{L}\,,\qquad
r_{2\,\rm circ}=-\frac{m}{L}\sqrt{2m^2+\sqrt{L^4+4m^4}}\,,
\eeq
for stable ($r_{1\,\rm circ}$) and unstable ($r_{2\,\rm circ}$) orbits, respectively, with $L$ necessarily negative (see also Fig. \ref{fig:Veff}).
The solution $r_{1\,\rm circ}$ has associated energy $E=1$ and vanishing $U^\phi$ component, so that it actually describes the dust particles which are at rest, as discussed above.
The solution $r_{2\,\rm circ}$ instead represents a true circular geodesic with four-velocity
\beq
U_{\rm circ}=\frac{r^4+m^4}{r^2\sqrt{r^4-m^4}}\left[\partial_t-\frac{m^2r}{r^4+m^4}\partial_\phi\right]\,.
\eeq
The associated energy and angular momentum are given by
\beq
E_{\rm circ}=\frac{r^2}{\sqrt{r^4-m^4}}\,,\qquad
L_{\rm circ}=-\frac{2m^2}{r}E_{\rm circ}\,.
\eeq
The angular velocity $\frac{d\phi}{dt}=-\frac{m^2 r}{r^4+m^4}$ tends to the constant value $-\frac{1}{2m}$ as the CTC boundary $r_{\rm ctc}=m$ is approached.
$U_{\rm circ}$ then becomes a null vector there (like $\partial_\phi$ on the same surface does).

\subsubsection{Helical-like motion around the symmetry axis}

\begin{figure}\[
\begin{array}{cc}
\includegraphics[scale=.20]{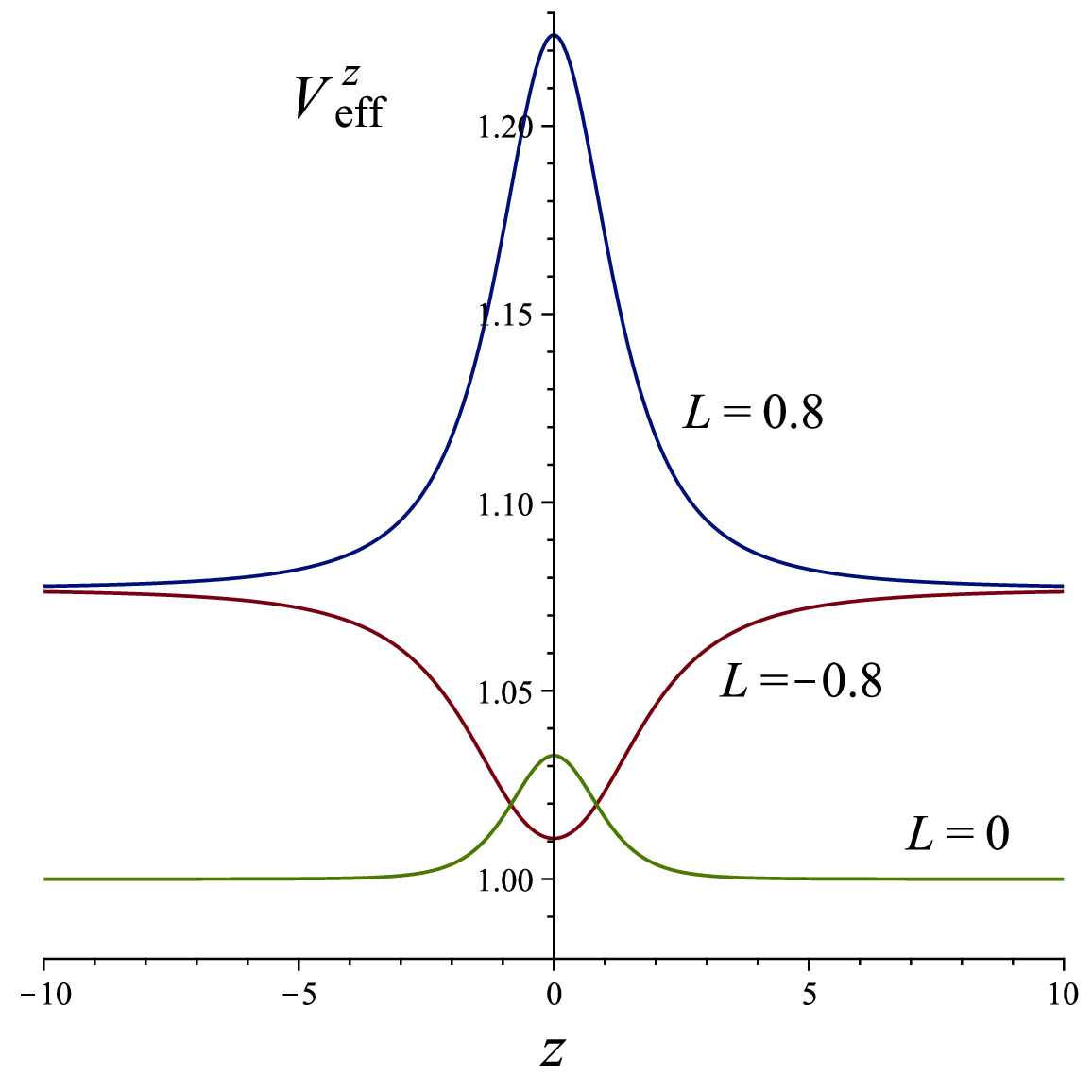}&
\includegraphics[scale=.20]{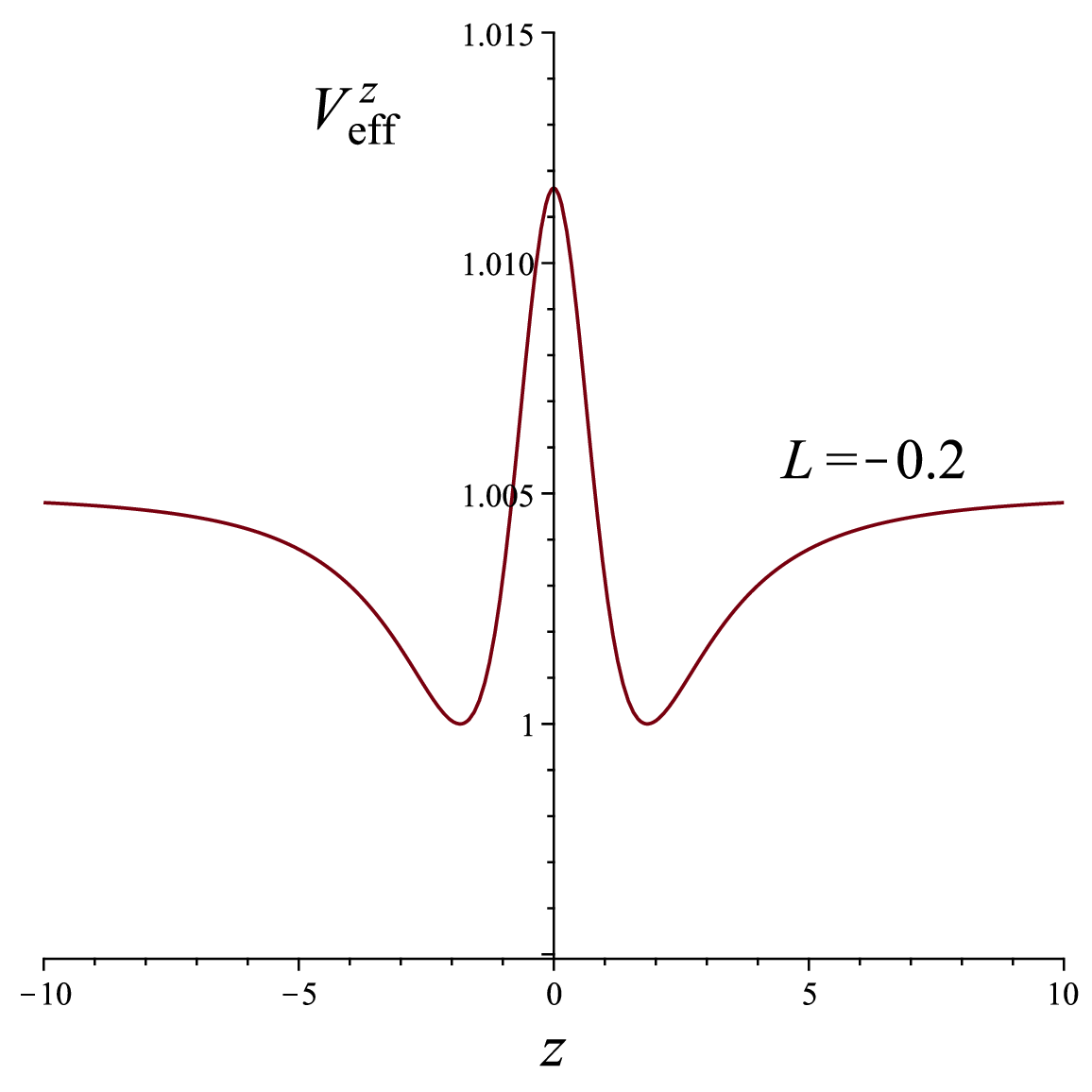}\cr
(a) & (b) \cr
\end{array}
\]
\caption{
\label{fig:Veffz}
Typical behavior of the (positive branch of the) effective potential for motion parallel to the symmetry axis as a function of $z$ for $m=1$, $r=2$ (implying that the orbit does not cross the CTC boundary) and different values of the particle's angular momentum parameter $L = [-0.8,0,0.8]$ (panel (a)), and $L = -0.2$ (panel (b)).
}
\end{figure}

Let us consider the helical-like motion around the symmetry axis, i.e., on the cylindrical surface $r=$ const.
One can introduce an effective potential also in this case, such that 
\beq
 g_{zz}(U^z)^2=\frac{ g_{\phi\phi}}{r^2}(E-V_{\rm eff}^{z\,+})(E-V_{\rm eff}^{z\,-})\,,
\eeq
with
\beq
g_{zz}V_{\rm eff}^{z\,\pm}=-g_{tz}L\pm r\sqrt{L^2+g_{zz}}\,.
\eeq
In the limit $z\to\pm\infty$ the positive branch of the potential goes to the value $\sqrt{r^2+L^2}/r$.
Turning points for $z$-motion are the real roots of the equation $U^z=0$, which can be solved as
\beq
(r^2+z^2)^{3/2}=-\frac{m^2r^2E}{L\pm r\sqrt{E^2-1}}\,.
\eeq
For instance, for $L=0$ there are two turning points at $z_\pm^*=\pm\left[\left(\frac{m^2rE}{\sqrt{E^2-1}}\right)^{2/3}-r^2\right]^{1/2}$.
The behavior of the (positive branch of the) effective potential is shown in Fig. \ref{fig:Veffz} for different values of the particle's angular momentum parameter.
Incoming particles endowed with enough energy generally cross the symmetry plane.
Low energy particles instead only approach the symmetry plane, and are forced to come back (see Fig. \ref{fig:Veffz} (a)). 
There also exist conditions allowing for confined motion around $z=0$ or even two opposite values of $z\not=0$, if the particle starts moving close to the symmetry plane with negative values of the angular momentum parameter (see Fig. \ref{fig:Veffz} (a) and (b), respectively).  

One can check that the condition to reach the CTC region writes $\coth\beta=1+{2\xi}/{\hat r}$ as before, but now $\hat r$ is a fixed parameter (for motion along $z$), which becomes 1 on the symmetry plane.

Circular geodesics on a $z=$ const. ($z\not=0$) hyperplane exist only for $z_\pm^{\rm circ}=\pm r/\sqrt{2}$.
However, the latter condition implies the vanishing of the angular velocity (with energy parameter $E=1$ and angular momentum parameter $L= -\frac{2\sqrt{6}m^2}{9r}$), so that particles actually are at rest there.
This is in agreement with the condition $\partial_zV_{\rm eff}^{z\,\pm}=0$, which gives  
\beq
z_\pm^{\rm circ}=\pm\left[\frac{(mrL)^{4/3}}{L^2}-r^2\right]^{1/2}\,,
\eeq
with $0<|L|\leq m^2/r$ ($L<0$, see Fig. \ref{fig:Veffz}), once $L$ is replaced by the value specified above.

\section{Non-geodesic motion under a friction force}

Let us consider now the more realistic situation in which the particle  (with mass $\mu$) moves along non-geodesic orbits due the interaction with the background fluid.
The latter can be modeled by adding to the equations of motion a friction force of the type 
\beq
\label{f_fric}
f_{\rm (fric)}(U)^\alpha =-\sigma P(U)^{\alpha}{}_{\mu}T^{\mu \nu}U_{\nu}\,,
\eeq
where $\sigma$ is the cross section of the process, $P(U)^{ \alpha}{}_{ \nu}=\delta^{ \alpha}{}_{ \nu}+U^{ \alpha} U_{ \nu}$  projects orthogonally to $U$, and $T^{\mu \nu}$ is the energy-momentum tensor of the fluid given by Eq. \eqref{T_munu}. 
Explicitly, we have
\beq
f_{\rm (fric)}(U)^\alpha = -\sigma \rho_m  (U\cdot u)[u^\alpha +(U\cdot u) U^\alpha ] \,,
\eeq
and the overall minus sign is purely conventional.
The force \eqref{f_fric} can be considered as the general relativistic generalization of the Stokes force acting on a body which moves in a fluid \cite{Bini:2013es}, even if non-viscous.
It is of the same form as that first introduced by Poynting and Robertson \cite{poy1,poy2} to study the effect of the radiation pressure of the light emitted by a star on the motion of small bodies moving around it. Recently, the role of the Poynting-Robertson effect in the problem of scattering of particles by a radiation field has also been investigated (see, e.g., Refs. \cite{Bini:2008vk,Bini:2011zza,Bini:2014sua,Bini:2011zz,Bini:2014yca}).

The force  \eqref{f_fric} has been already adopted in different contexts: to study non-geodesic motion in the spacetime of self-gravitating fluids\footnote{It also couples the matter-energy distribution with the geometry.} \cite{Bini:2012ff,Bini:2013es,Bini:2014lwz}, and to discuss from a cosmological perspective how (test) fluids describing either ordinary or exotic matter and surrounding a non-rotating source may scatter particles \cite{Bini:2012zzd}. 

\begin{figure}\[
\begin{array}{cc}
\includegraphics[scale=.20]{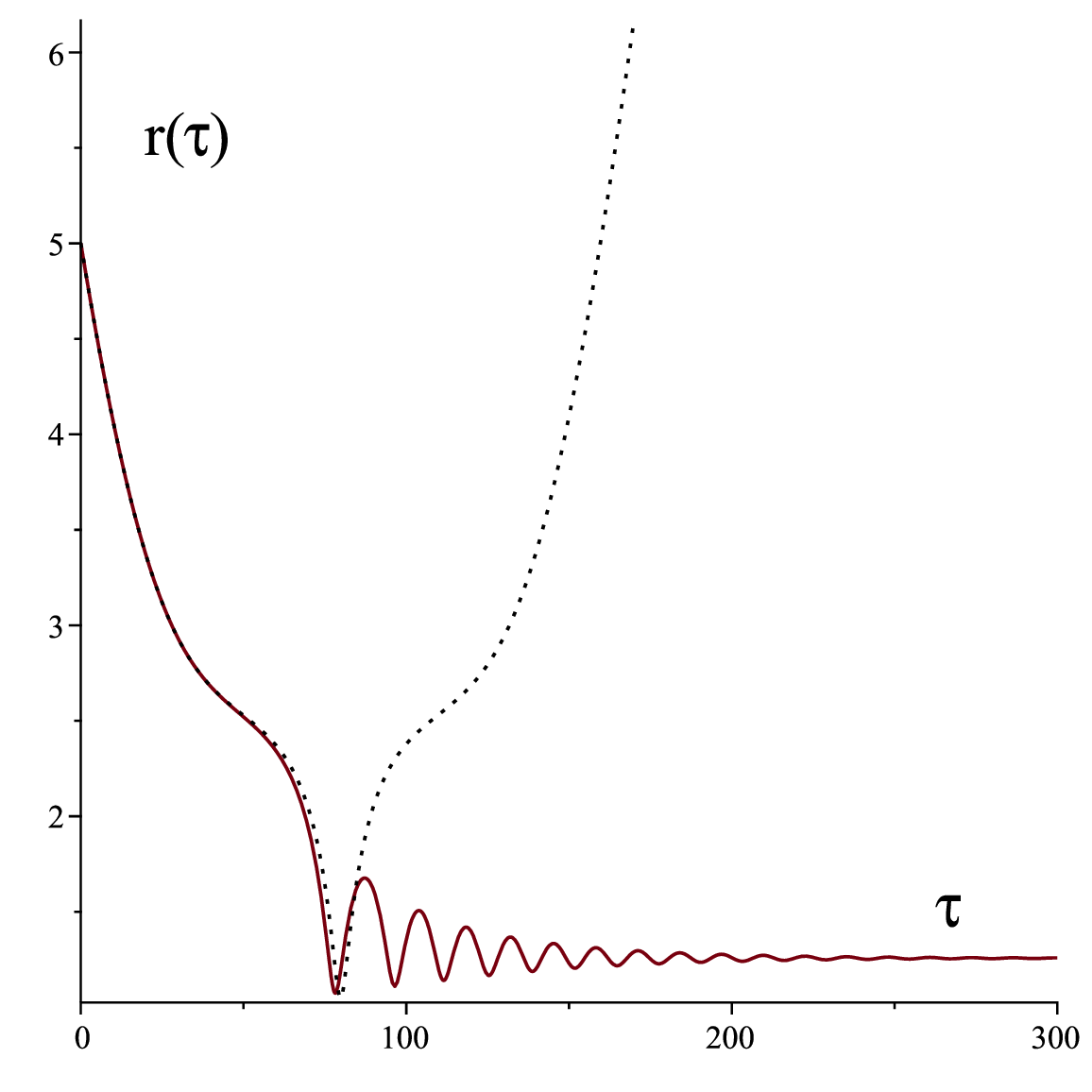}&
\includegraphics[scale=.20]{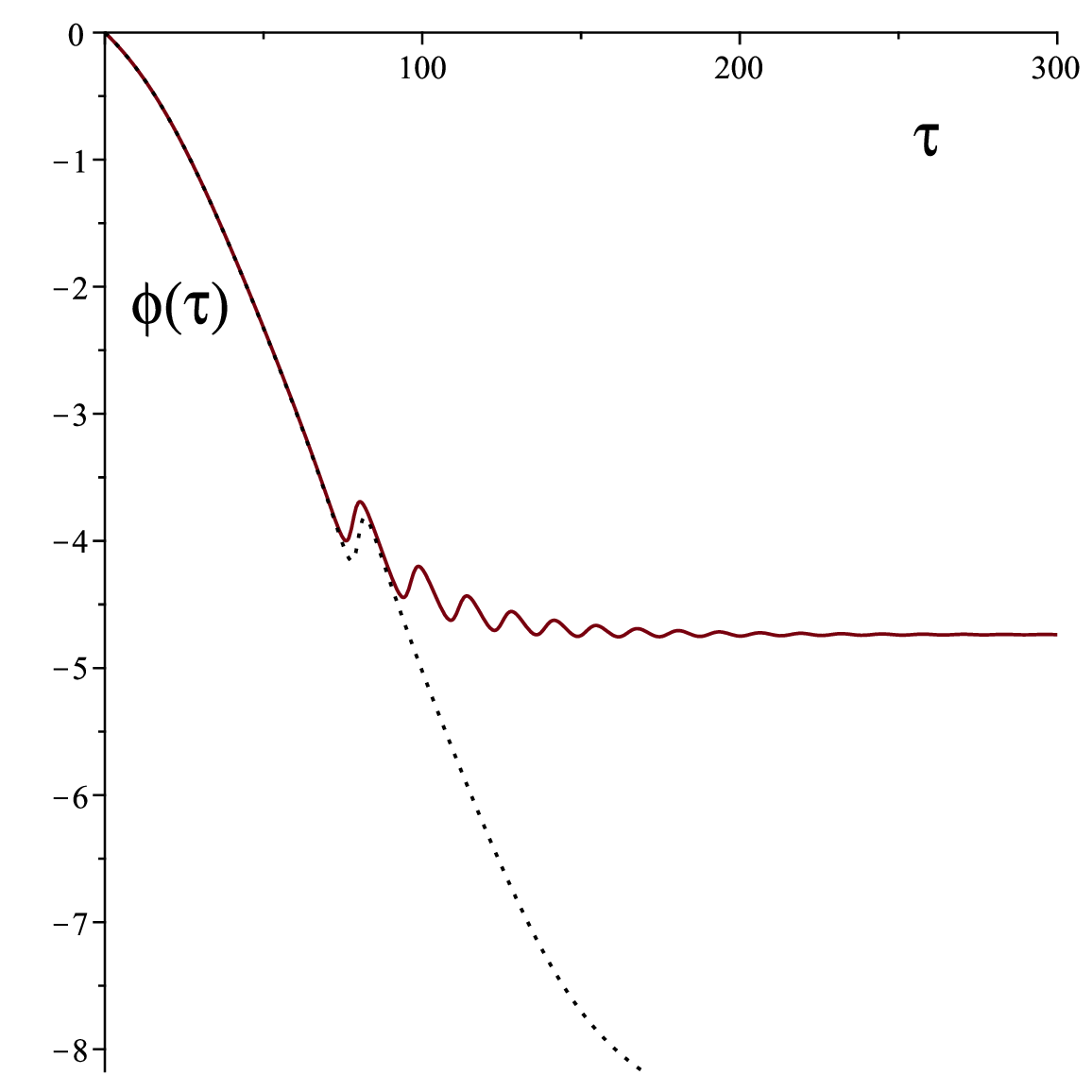}\cr
(a) & (b) \cr
\end{array}
\]
\[\begin{array}{c}
\includegraphics[scale=.20]{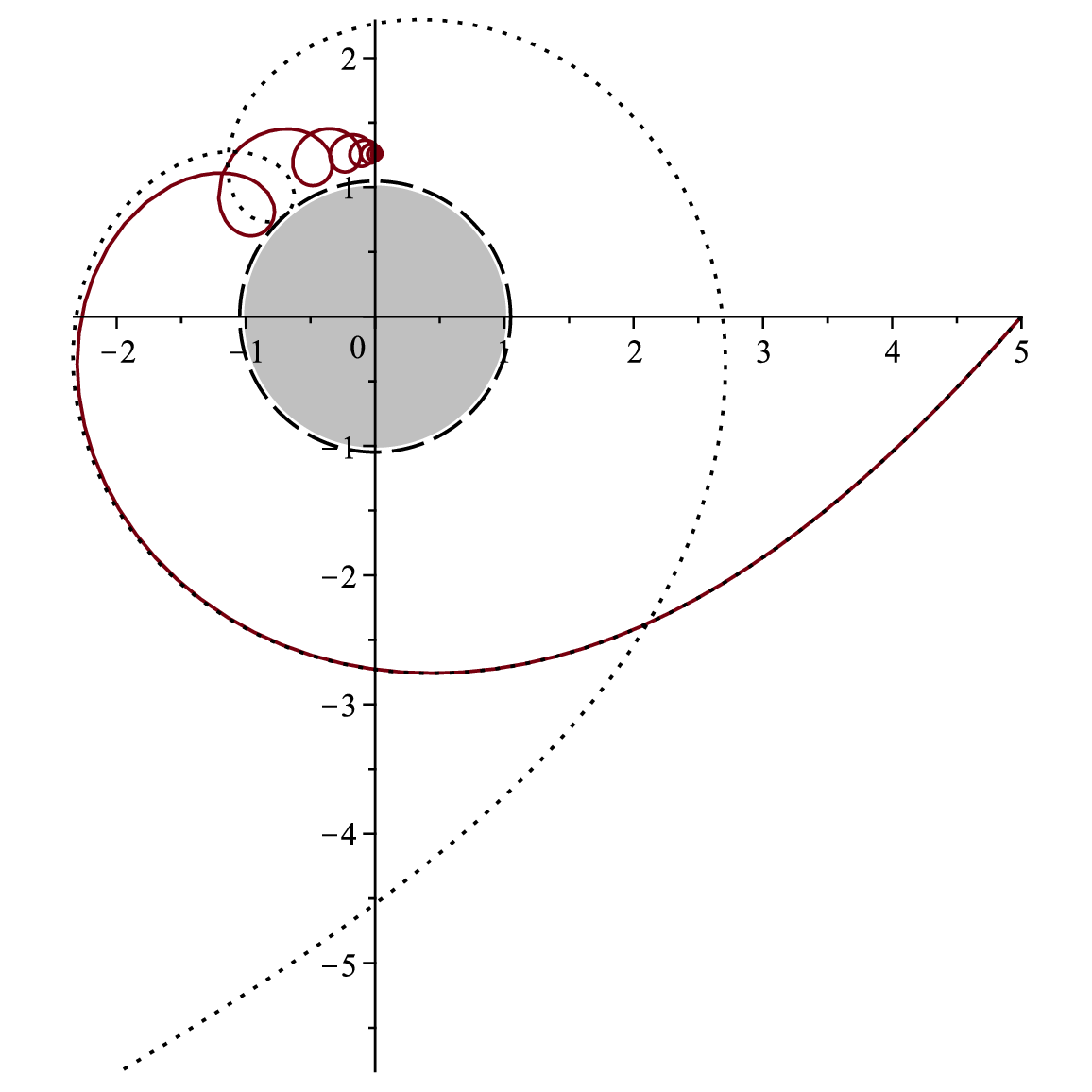}\cr
(c) \cr
\end{array}
\]
\caption{
\label{fig:9}
Example of numerical integration of the nongeodesic orbit due to the friction force in the symmetry plane $z=0$, corresponding to the same initial conditions as for the geodesic orbit of Fig. \ref{fig:equat} (b), also superposed to the plot with dotted style.
Here, we have chosen $\tilde \sigma=0.1$ in order to enhance the effect visually.
The orbit is no more hyperboliclike, but it stops at $r\approx1.2584$ and $\phi\approx-271.4451$ degrees because of the drag due to the interaction with the fluid.
}
\end{figure}

Our motivation here is to see whether the presence of a friction force can slow-down (rapidly enough) the motion, so preventing particles from reaching the pathological region containing CTCs. 
The equations of motions write as 
\beq
\label{moteqs}
\frac{d U^\alpha}{d\tau }=f_{\rm (grav)}(U)^\alpha+f_{\rm (fric)}(U)^\alpha\,,
\eeq
where
\bea
f_{\rm (grav)}(U)^\alpha&=& -\Gamma^\alpha{}_{\beta\sigma}U^\beta U^\sigma \,,\nonumber\\
f_{\rm (fric)}(U)^\alpha&=&-\tilde \sigma \kappa \rho_m  U_t [u^\alpha +U_t U^\alpha ]\,,
\eea
with our choice of coordinates (fluid at rest in the chosen coordinate system, $u^\alpha=\delta^\alpha_0$), and $\tilde \sigma=\frac{\sigma}{\kappa \mu}$.
Here $f_{\rm (grav)}(U)$ depends on $U$ but it is not orthogonal to $U$ itself, differently from $f_{\rm (fric)}(U)$ which is orthogonal to $U$ by definition.
Limiting our considerations to motion on the symmetry plane ($z=0$, $U^z=0$), for simplicity, the equations of motion \eqref{moteqs} then read 
\bea
\label{sys_PR}
\frac{d U^r}{d\tau }&=&- \tilde\sigma \frac{m^4e^{-\frac{m^4}{8r^4}}}{r^6}{\mathcal E}^2U^r  
+\frac{e^{-\frac{m^4}{8r^4}}}{r^2}(m^2 {\mathcal E}+r^3 U^\phi)U^\phi\nonumber\\
&+&\frac{m^4}{4r^5}(U^r)^2
\,,\nonumber\\
\frac{d U^\phi}{d\tau }&=&- \tilde\sigma \frac{m^4e^{-\frac{m^4}{8r^4}}}{r^6} {\mathcal E}^2 U^\phi
-\frac1{r^4}(m^2 {\mathcal E}+2r^3 U^\phi)U^r
\,,\nonumber\\
\frac{d {\mathcal E}}{d\tau }&=&- \tilde\sigma \frac{m^4e^{-\frac{m^4}{8r^4}}}{r^6}{\mathcal E}({\mathcal E}^2-1)\,,
\eea
where
\beq
{\mathcal E}=\frac1r(r U^t+m^2 U^\phi)\,,
\eeq
with the normalization condition for $U$, $U\cdot U=-1$, 
\beq
-{\mathcal E}^2+r^2(U^\phi)^2+e^{\frac{m^4}{8r^4}}(U^r)^2=-1\,.
\eeq
Here ${\mathcal E}$ reduces to the conserved energy $E$ along geodesic motion on the symmetry plane, see Eq. \eqref{EL_defs}.
As a typical feature $f_{\rm (grav)}(U)$ and $f_{\rm (fric)}(U)$ compete among themselves, and one can usually find  a \lq\lq suspended orbit" radius at which the particle can stay at rest, which is an equilibrium solution for the above system in the asymptotic regime.
An example of numerical integration of the orbits is shown in Fig. \ref{fig:9}.

One can look for solutions of the system \eqref{sys_PR} in the limit of small values of $\tilde \sigma$.
The linear-in-$\tilde \sigma$ solution is straightforward,  and implies oscillating behaviors around $r=r_0$ and $\phi=\phi_0$ with first-order modifications of the four-velocity components given by
\bea
U^\phi_1(\tau)&=& -\frac{m^2}{\Omega_0 r_0^4}  U^r_1(0)\sin \Omega_0 \tau +U^\phi_1(0)\cos \Omega_0 \tau\,,\nonumber\\
U^r_1(\tau)&=&U^r_1(0)\cos \Omega_0 \tau +\frac{\Omega_0 r_0^4}{m^2}U^\phi_1(0)\sin\Omega_0 \tau\,,
\eea 
where $\Omega_0=\frac{m^2}{r_0^3}e^{-\frac{m^4}{16 r_0^4}}$, and ${\mathcal E}=1+O(\tilde\sigma^2)$.

\section{Concluding remarks}

For a fluid-sourced spacetime the possibility to study gravitational effects as well as particle-particle interactions should be put beside by gravitationally induced collective motion of the fluid particles. 
This is the case of accreting matter and radiation in the spacetime region around compact objects, for example.
In a realistic scenario the thermodynamic properties of the fluid are expected to strongly influence the motion of particles inside it.

The simplest approximation consists in considering a test fluid superposed to the (fixed) gravitational field of the compact object, e.g., a distribution of collisionless dust surrounding a Schwarzschild black hole \cite{Bini:2016tqz}.
The interaction between test particles moving in the background spacetime and the surrounding dust has been modeled there by adding to the equations of motion a friction force built with the stress-energy tensor of the dust, which is responsible for a loss of both energy and angular momentum, causing the particle trajectory to be deflected with respect to the corresponding geodesic path.

In the case of a spacetime sourced by a fluid, instead, it is the matter-energy content of the fluid itself which causes the curvature of the spacetime (a completely different situation with respect to the test fluid case).
The interaction of test particles with the background fluid in a first approximation (i.e., by neglecting particle's backreaction on the background geometry) is thus described by (timelike) geodesics.
While test particle dynamics around black holes is well studied in the literature, the same is not true in the case of fluid-sourced spacetimes. In fact, the latter are in general plagued by the presence of some pathological behavior, like the unphysical behavior of thermodinamical quantities, and the existence of spacetime regions where causality as well as energy conditions are violated, besides singularities.

In this paper we have investigated test particle motion in the spacetime of a rotating dust, known as Bonnor's spacetime.
The dust particles form a congruence which is geodesic, has vanishing expansion and shear, but nonzero vorticity.
The spacetime has a singularity at the origin of the coordinates, which is  the result of the nonlinearities of the gravitational field due the kinematical properties of the fluid; in particular, its vorticity field, which diverges there.
In addition, close to the singularity there exists a toroidal region where CTCs are allowed, which is usually discarded as unphysical. The boundary of such a region reduces to a circle on the symmetry plane.

We have numerically integrated the geodesic equations with initial conditions chosen such that particles are released from a fixed space point and directed towards the singularity. The study of the effective potential for radial motion on the symmetry plane has shown that for most of the allowed values of particle's energy and angular momentum there is a barrier typically preventing particles to reach the CTC boundary (and the singularity). We have identified indeed conditions on the space of the parameters such that these orbits cannot enter the CTC region. Moreover, to make this circumstance more favored we have introduced a friction  force (modeled \'a la Poynting-Robertson \cite{poy1,poy2}) which drags particles after their releasing, eventually stopping them before they reach the pathological region containing CTCs. 
Noticeably, we have found conditions for the existence of circular orbits (which are but all unstable), as well as bound motion between a minimum and a maximum radius.
The latter are well known features of motion around compact objects, but a priori not expected in this case.

\section*{Acknowledgements}

D.B. acknowledges sponsorship of the Italian Gruppo Nazionale per la
Fisica Matematica (GNFM) of the Istituto Nazionale di
Alta Matematica (INDAM). M.L.R. acknowledges the contribution of  the local research  project ``Modelli gravitazionali per lo studio dell'universo'' (2022) - Dipartimento di Matematica ``G. Peano,'' Universit\`a degli Studi di Torino and of the Gruppo Nazionale per la Fisica Matematica (GNFM).
D.A. is supported by the Icelandic Research Fund under grant 228952-052.

\appendix

\section{Separable solutions to Eq. \eqref{GFequation2} \label{Solgenapp}}

In this Appendix we review the construction of separable solutions to Eq. \eqref{GFequation2}, namely
\beq
\partial_{zz}A+\partial_{rr}A-\frac{1}{r }\partial_r A=0\,,
\eeq
with $A={\mathcal R}(r){\mathcal Z}(z)$, obtained in Refs. \cite{Cooperstock:2005qw,Balasin:2006cg}.
The above ansatz gives
\bea
    && {\mathcal Z}_{zz}=k{\mathcal Z}\,,\nonumber\\
&& {\mathcal R}_{rr}-\frac1r {\mathcal R}_r+k {\mathcal R}=0\,,
\eea
with the separation constant $k\in R$. Imposing the reflection symmetry property $A(r,z)=A(r,-z)$ 
leads to two possibilities, according to positive or negative values of $k$. 

\begin{enumerate}
\item One assumes a positive value for $k=u^2$ (with $u\in R_0^+$) and the fall off behavior at infinity can be achieved by employing the non-smooth modes ${\mathcal Z}(k,z)=e^{-\sqrt{k}|z|}=e^{-u |z|}$, which satisfy Eq. \eqref{GFequation2} for $z>0$ only.
Consequently, because of the lack of regularity in the metric at $z=0$,  there are sources localized at $z=0$. To find an explicit  solution in this case  
one starts from 
\beq
    \tilde{U}(u,z)= \int^\infty_0  A(r,z) J_1(ru) dr\,,
\eeq
where $J_m(ru)$ is the Bessel function of the order $m$. Then,  
\bea
    \partial_{zz} \tilde{U}&=&\int^\infty_0 \partial_{zz} A(r,z) \, J_2(ru) dr  \nonumber\\
    &=& -\int^\infty_0 \left( \partial_{rr}A-\frac{1}{r}\partial_r A\right) J_1(ru) dr\,.
\eea 
Double integration by part gives
\beq
    \partial_{zz} \tilde{U}=u^2 \int^\infty_0 A \,  J_1(ru) dr = u^2 \tilde{U}\,.
\eeq
This implies that the required form is $\tilde{U}(u,z)= \alpha_+(u) e^{-u  z }+\alpha_- (u)e^{u z }$.

The asymptotic  condition 
\beq
    \lim_{z\rightarrow\pm \infty} A=0\,,
\eeq
implies $\alpha_- (u)=0$ for $z>0$ and $\alpha_+ (u)=0$ for $z<0$. Summarizing 
\bea
     \tilde{U}(u,z)&=&\alpha(u)e^{-u|z|}=\int^\infty_0  A(r,z) J_1(ru) dr\,, 
\eea
with $\alpha(u)$ arbitrary.
To obtain the final expression for the function $A$ now it suffices to invert the above relation by using the Hankel transform. The result is
\beq
    A(r,z)= \int^\infty_0 r u \alpha(u) e^{-u|z|} J_1(ru) du\,,
\eeq
due to the arbitrariness of $\alpha(u)$ we can redefine a new function using $\tilde{A}(u)=u \alpha(u)$ such that
\beq
    \psi(r,z)=\int^\infty_0 r e^{-\lambda |z|}\tilde{A}(\lambda) J_1(\lambda r) d\lambda\,. \label{psi}
\eeq

\item Another possibility is  ${\mathcal Z}(k,z)=\cosh{(\sqrt{k}z)}$. The parameter $k$ cannot be positive then, otherwise these modes diverge exponentially for $|z|\to\infty$, which is an unphysical behavior. In this case we assume $k=-\lambda^2$ with $\lambda\in R_0^+$ yielding the modes ${\cal Z}(\lambda,z)=\cos{(\lambda z)}$.
The solution can be found with the same procedure as in the first case. The result is given by 
\bea
\label{eq:ax8}  
A(r,z)&=&\int\limits_0^\infty d\lambda \cos{(\lambda z)}(r\lambda)\left[a(\lambda)K_1(\lambda r)\right.\nonumber\\
&&\left.+b(\lambda)I_1(\lambda r)\right]\,.
\eea
The functions $I_1$ and $K_1$ are modified Bessel functions of the first and second kind, respectively. Since $I_1$ blows up exponentially for large values of $r$ we set $b(\lambda)=0$. The function $K_1$ falls off exponentially for large values of $r$ and diverges like $1/r$ near $r=0$. However, this divergence is compensated by a linear prefactor, so the integrand is well defined for any sufficiently regular $a(\lambda)$.
We can write   the result in a simpler form. We perform first a Fourier transformation,
\beq
  \label{eq:ax25}
  a(\lambda)=\frac{2}{\pi}\int\limits_0^\infty d x \,C(x)\cos{(\lambda x)}\,,
\eeq
where $a(\lambda)$ is determined in terms of a (Fourier) transformed spectral density $C(x)$.
Using the  property  
\begin{equation}
  \label{eq:ax23}
  \int\limits_0^\infty d x x K_1(x) \cos \left(\frac{c\, x}{r}\right)  = \frac{\pi}{2}\frac{r^3}{(c^2+r^2)^{3/2}}\,,
\end{equation}
the form of $A$ becomes 
\bea
\label{Gensol}
  A(r,z)  
&=& \frac{1}{2}\int\limits_{-\infty}^\infty d\zeta  \frac{C(\zeta)r^2}{[(z+\zeta)^2+r^2]^{3/2}}\,.
\eea
\end{enumerate}

\end{document}